\documentclass[aps,prb,twocolumn,amsmath,amssymb,showpacs,superscriptaddress]{revtex4}

\usepackage{graphicx}
\usepackage{dcolumn}
\usepackage{bm}

\begin{document}

\title{Surface plasmon polaritons and surface phonon polaritons \\
on metallic and semiconducting spheres: \\ Exact and semiclassical
descriptions}

%\title{Surface polaritons on metallic and semiconducting spheres}

\author{St\'ephane Ancey}
\email{ancey@univ-corse.fr}
\affiliation{ UMR CNRS 6134 SPE, Equipe Ondes et Acoustique, \\
Universit\'e de Corse, Facult\'e des Sciences, Bo{\^\i}te Postale
52, 20250 Corte, France}

\author{Yves D\'ecanini}
\email{decanini@univ-corse.fr}
\affiliation{ UMR CNRS 6134 SPE,
Equipe Physique Semi-Classique (et) de la Mati\`ere Condens\'ee,
\\ Universit\'e de Corse, Facult\'e des Sciences, Bo{\^\i}te
Postale 52, 20250 Corte, France}

\author{Antoine Folacci}
\email{folacci@univ-corse.fr}
\affiliation{ UMR CNRS 6134 SPE,
Equipe Physique Semi-Classique (et) de la Mati\`ere Condens\'ee,
\\ Universit\'e de Corse, Facult\'e des Sciences, Bo{\^\i}te
Postale 52, 20250 Corte, France}

\author{Paul Gabrielli}
\email{gabrieli@univ-corse.fr}
\affiliation{ UMR CNRS 6134 SPE,
Equipe Ondes et Acoustique,
\\ Universit\'e de Corse, Facult\'e des Sciences, Bo{\^\i}te
Postale 52, 20250 Corte, France}

\date{\today}

\begin{abstract}

We study the interaction of an electromagnetic field with a
non-absorbing or absorbing dispersive sphere in the framework of
complex angular momentum techniques. We assume that the dielectric
function of the sphere presents a Drude-like behavior or an ionic
crystal behavior modelling metallic and semiconducting materials. We
more particularly emphasize and interpret the modifications induced
in the resonance spectrum by absorption. We prove that ``resonant
surface polariton modes" are generated by a unique surface wave,
i.e., a surface (plasmon or phonon) polariton, propagating close to
the sphere surface. This surface polariton corresponds to a
particular Regge pole of the electric part (TM) of the $S$ matrix of
the sphere. From the associated Regge trajectory we can construct
semiclassically the spectrum of the complex frequencies of the
resonant surface polariton modes which can be considered as
Breit-Wigner-type resonances. Furthermore, by taking into account
the Stokes phenomenon, we derive an asymptotic expression for the
position in the complex angular momentum plane of the surface
polariton Regge pole. We then describe semiclassically the surface
polariton and provide analytical expressions for its dispersion
relation and its damping in the non-absorbing and absorbing cases.
In these analytic expressions, we more particularly exhibit
well-isolated terms directly linked to absorption. Finally, we
explain why the photon-sphere system can be considered as an
artificial atom (a ``plasmonic atom" or ``phononic atom") and we
briefly discuss the implication of our results in the context of the
Casimir effect.

\end{abstract}


\maketitle

\section{Introduction}

In a recent article \cite{ADFG_1}, we introduced the complex angular
momentum (CAM) method in the context of scattering of
electromagnetic waves from non-absorbing metallic and semiconducting
cylinders. This allowed us to completely describe the resonant
aspects of the problem in the frequency range where the dielectric
function is negative. In the present article, we shall extend this
analysis to non-absorbing as well as absorbing metallic and
semiconducting spheres. For non-absorbing media, the transition from
two dimensions to three dimensions induces some additional technical
difficulties as well as the appearance of curvature correction terms
in some analytic formulas describing surface polaritons (SP's) while
the physical interpretations remain quasi-identical in form. As a
consequence, we will pass quickly on certain aspects lengthily
analyzed in Ref.~\onlinecite{ADFG_1}. By contrast, for absorbing
media, different physical phenomenons occur. They can be interpreted
from new correction terms (mainly due to the imaginary part of the
complex dielectric constant) in the analytical expression describing
the damping of SP's.

The theory of the resonant surface polariton modes (RSPM's)
supported by metallic and semiconducting spheres has been already
studied in numerous works (see, for example,
Refs.~\onlinecite{FuchsKliewer1968,EngRup68a,EngRup68b,EngRup68c,RuppinR82,Martinos85,
Ruppin98,Sernelius01,DungETAL2001,InglesfieldETAL2004,PitarkeETAL2007}
and references therein) and is currently the subject of a renewed
interest in the fields of nanotechnologies and plasmonics (for a
recent comprehensive review on this subject, we refer to
Ref.~\onlinecite{Atwater2007}). In the present article, by using CAM
techniques\cite{New82,Nus92,Grandy} in connection with modern
aspects of asymptotics
\cite{Dingle73,Berry89,BerryHowls90,SegurTL91}, we shall look
further into this subject. The CAM method has been extensively used
in physics (we refer to the Introduction of Ref.~\onlinecite{ADFG_1}
for a description of this method and for a short bibliography). We
have recently introduced it in the context of electromagnetism of
dispersive media (see Refs.~\onlinecite{ADFG_1,ADFG_2,ADFG_4}).
Here, by using this method, we shall provide a clear physical
explanation for the excitation mechanism of the RSPM's of the sphere
as well as a simple mathematical description of the unique surface
wave, i.e. of the so-called surface polariton (SP), that generates
them.

Our paper is organized as follows. Section II is devoted to the
exact theory: we introduce our notations, we provide the expression
of the $S$ matrix of the system and we then discuss the resonant
aspects of the problem. In Sec. III, by using CAM techniques, we
establish the connection between the SP propagating close to the
surface sphere and the associated RSPM's. In Sec. IV, we describe
semiclassically the SP by providing analytic expressions for its
dispersion relation and its damping. Finally, in Sec. V, we conclude
our article by explaining why the photon-sphere system can be viewed
as an artificial atom (a ``plasmonic atom" or ``phononic atom") and
by briefly discussing the implication of our results in the context
of the Casimir effect.

\section{Exact $S$ matrix and scattering resonances}

\subsection{General theory}

From now on, we consider the interaction of an electromagnetic field
with a metallic or semiconducting sphere with radius $a$ which is
embedded in a host medium of infinite extent. In the usual spherical
coordinate system $(r ,\theta ,\varphi)$ the sphere occupies a
region corresponding to the range $0 \le r < a$ (region II) while
the host medium corresponds to the range $ r > a$ (region I). In the
following, we implicitly assume a time dependence in $\exp(-i\omega
t)$ for the electromagnetic field and we denote by $\epsilon_c
(\omega)$ the frequency-dependent dielectric function of the sphere
and by $\epsilon_h$ the constant dielectric function of the host
medium. Furthermore, we shall use the wave numbers
\begin{equation}
k^{\mathrm {I}}(\omega)=\left(\frac{\omega }{c} \right)
\sqrt{\epsilon_h} \quad \mathrm{and} \quad k^{\mathrm
{II}}(\omega)=\left(\frac{\omega }{c} \right) \sqrt{ \epsilon_c
(\omega)}
\end{equation}
in order to describe wave propagation in regions I and II (here $c$
is the velocity of light in vacuum).

As far as the dielectric function of the sphere is concerned, we
assume it presents a Drude-like behavior
\cite{AshcroftMermin,MarkFox}
\begin{equation}
\epsilon_c (\omega)= \epsilon_\infty \left( 1-
\frac{\omega_p^2}{\omega ^2 + i\gamma \omega_p \omega} \right),
\label{PermDrude}
\end{equation}
or an ionic crystal behavior
\cite{FuchsKliewer1968,AshcroftMermin,MarkFox}
\begin{equation}
\epsilon_c (\omega)=  \epsilon_\infty \left( \frac{\omega_L^2 -
\omega^2 -i\gamma \omega_T \omega}{\omega_T^2 -  \omega^2 -i\gamma
\omega_T \omega} \right). \label{PermCristIon}
\end{equation}
In both cases, $\epsilon_\infty$ is the high-frequency limit of the
dielectric function and $\gamma$ is a phenomenological damping
factor. In Eq.~(\ref{PermDrude}), $\omega_p$ is the plasma
frequency. In Eq.~(\ref{PermCristIon}), $\omega_T$ and $\omega_L$
respectively denote the transverse-optical-phonon frequency and the
longitudinal-optical-phonon frequency. In the first case, SP's
follow from the coupling of the electromagnetic wave with the plasma
wave and are usually called surface plasmon polaritons. In the
second one, SP's follow from the coupling of the electromagnetic
wave with the longitudinal and transverse acoustic waves and are
usually called surface phonon polaritons. Eq.~(\ref{PermDrude}) can
be used to describe the dielectric behavior of certain metals and
semiconductors (Si, Ge, InSb, $\dots$ ) while
Eq.~(\ref{PermCristIon}) can be used to investigate the optical
properties of other semiconductors such as GaAs.

In the following, we shall often consider separately the real and
imaginary parts of the dielectric function. We can write
\begin{equation}
\epsilon_c (\omega)=  \epsilon'_c (\omega) + i \epsilon''_c (\omega)
\label{FuncDiel_RetI}
\end{equation}
with
\begin{subequations}
\begin{eqnarray} \label{FuncDiel_RetI_Met}
& & \epsilon'_c (\omega) =  \epsilon_\infty \left[ 1-
\frac{\omega_p^2}{\omega ^2 + (\gamma \omega_p)^2} \right]     \label{FuncDiel_RetI_Met_a}\\
& & \epsilon''_c (\omega) = \epsilon_\infty \left[ \frac{\gamma
\omega_p^2 (\omega_p/\omega)}{\omega ^2 + (\gamma \omega_p)^2}
\right] \label{FuncDiel_RetI_Met_b}
\end{eqnarray}
\end{subequations} for the Drude-like behavior and
\begin{subequations}
\begin{eqnarray} \label{FuncDiel_RetI_SC}
& & \epsilon'_c (\omega) =  \epsilon_\infty \left[ \frac{(\omega_L^2
- \omega^2)(\omega_T^2 - \omega^2) + (\gamma \omega_T
\omega)^2}{(\omega_T^2 - \omega^2)^2 + (\gamma \omega_T
\omega)^2} \right]   \label{FuncDiel_RetI_SC_a} \\
& & \epsilon''_c (\omega) =  \epsilon_\infty \left[ \frac{ \gamma
\omega_T \omega(\omega_L^2 - \omega_T^2 )  }{(\omega_T^2 -
\omega^2)^2 + (\gamma \omega_T \omega)^2} \right]
\label{FuncDiel_RetI_SC_b}
\end{eqnarray}
\end{subequations}
for the ionic crystal behavior. It is important to note that the
phenomenological damping factor $\gamma$ is always smaller than the
other parameters involved in the expression of $\epsilon_c
(\omega)$. As a consequence, we can always consider that
$|\epsilon''_c (\omega)| \ll |\epsilon'_c (\omega)|$.

The $S$ matrix of the sphere is of fundamental importance because it
contains all the information about the interaction of the
electromagnetic field with the sphere. It can be obtained from
Maxwell's equations and usual continuity conditions for the electric
and magnetic fields at the interface between regions I and II
\cite{Stratton,Nus92,Grandy}. For our problem, the elements of the
electric part (TM) of the $S$ matrix are given by
\begin{equation}
S_\ell^E(\omega)=  1 -2a_\ell^E(\omega) \label{SE1}
\end{equation}
with
\begin{equation}
a_\ell^E(\omega)= \frac{C^E_\ell(\omega)}{D^E_\ell(\omega)}
\label{SE2}
\end{equation}
where $C^E_\ell(\omega)$ and $D^E_\ell(\omega)$ are two $2\times 2$
determinants which are explicitly given by
\begin{subequations}\label{SE3ab}
\begin{eqnarray}
C_{\ell }^E(\omega ) &=&k^{\text{\textrm{II}}}\left( \omega \right)
\psi _{\ell }\left[ k^{\text{\textrm{II}}}\left( \omega \right)
a\right] \psi _{\ell }^{\prime }\left[ k^{\text{\textrm{I}}}\left(
\omega \right) a\right] \nonumber \\
& & \,\, -k^{\text{\textrm{I}}}\left( \omega \right) \psi _{\ell
}\left[ k^{\text{\textrm{I}}}\left( \omega \right) a\right] \psi
_{\ell }^{\prime }\left[ k^{\text{\textrm{II}}}\left( \omega
\right) a\right]  \label{SE3a} \\
D_{\ell }^E(\omega ) &=&k^{\text{\textrm{II}}}\left( \omega \right)
 \psi _{\ell }\left[ k^{\text{\textrm{II}}}\left( \omega \right)
a\right] \zeta _{\ell }^{(1)^{\prime }}\left[ k^{\text{\textrm{I}}%
}\left( \omega \right) a\right] \nonumber \\
& & \,\, -k^{\text{\textrm{I}}}\left( \omega \right) \zeta _{\ell
}^{(1)}\left[ k^{\text{\textrm{I}}}\left( \omega
\right) a\right] \psi _{\ell }^{\prime }\left[ k^{\text{\textrm{II}}%
}\left( \omega \right) a\right] \label{SE3b}
\end{eqnarray}%
\end{subequations}
while the elements of its magnetic part (TE) are given by
\begin{equation}
S_\ell^M(\omega)=  1-2a_\ell^M(\omega) \label{SM1}
\end{equation}
with
\begin{equation}
a_\ell^M(\omega)= \frac{C^M_\ell(\omega)}{D^M_\ell(\omega)}
\label{SM2}
\end{equation}
where $C^M_\ell(\omega)$ and $D^M_\ell(\omega)$ are also two
$2\times 2$ determinants which are now explicitly given by
\begin{subequations}\label{SM3ab}
\begin{eqnarray}
C_{\ell }^M(\omega ) &=&k^{\text{\textrm{II}}}\left( \omega \right)
\psi _{\ell }\left[ k^{\text{\textrm{I}}}\left( \omega \right)
a\right] \psi _{\ell }^{\prime }\left[ k^{\text{\textrm{II}}}\left(
\omega \right) a\right] \nonumber \\
& & \,\, -k^{\text{\textrm{I}}}\left( \omega \right) \psi _{\ell
}\left[ k^{\text{\textrm{II}}}\left( \omega \right) a\right] \psi
_{\ell }^{\prime }\left[ k^{\text{\textrm{I}}}\left( \omega
\right) a\right]  \label{SM3a}\\
D_{\ell }^M(\omega ) &=&k^{\text{\textrm{II}}}\left( \omega \right)
\zeta _{\ell }^{(1)}\left[ k^{\text{\textrm{I}}}\left( \omega
\right) a\right] \psi _{\ell }^{\prime }\left[ k^{\text{\textrm{II}}
}\left( \omega \right) a\right] \nonumber \\
& & \,\, -k^{\text{\textrm{I}}}\left( \omega \right) \psi _{\ell
}\left[ k^{\text{\textrm{II}}}\left( \omega \right) a\right] \zeta
_{\ell }^{(1)^{\prime }}\left[ k^{\text{\textrm{I}}  }\left( \omega
\right) a\right]. \nonumber \\
& &    \label{SM3b}
\end{eqnarray}
\end{subequations}
In Eqs.~(\ref{SE3ab}) and (\ref{SM3ab}), we use the Ricatti-Bessel
functions $\psi_\ell(z)$ and $\zeta _{\ell }^{(1)}(z)$ which are
linked to the spherical Bessel functions $j_\ell(z)$ and
$h^{(1)}_\ell(z)$ by $\psi_\ell(z)=zj_\ell(z)$ and $\zeta _{\ell
}^{(1)}(z)=zh^{(1)}_\ell(z)$ (see Ref.~\onlinecite{AS65}).

From the $S$ matrix elements, we can in particular construct the
scattering cross section $\sigma_\mathrm{sca}$ and the absorption
cross section $\sigma_\mathrm{abs}$ of the
sphere\cite{Nus92,Grandy}. They can be expressed in terms of the
coefficients $a_\ell^E(\omega)$ and $a_\ell^M(\omega)$ and they are
given by
\begin{equation}\label{crosssection_sca}
\sigma_\mathrm{sca} (\omega) =
\frac{2\pi}{\left[k^{\text{\textrm{I}}}\left( \omega
\right)\right]^2}\sum_{\ell=1}^{\infty}
(2\ell+1)[|a_\ell^E(\omega)|^2+|a_\ell^M(\omega)|^2]
\end{equation}
and
\begin{eqnarray}\label{crosssection_abs}
& & \sigma_\mathrm{abs} (\omega) =
\frac{2\pi}{\left[k^{\text{\textrm{I}}}\left( \omega
\right)\right]^2}\sum_{\ell=1}^{\infty} (2\ell+1)\left\{
[\mathrm{Re} \, a_\ell^E(\omega)
-|a_\ell^E(\omega)|^2] \right. \nonumber \\
& & \qquad\qquad\qquad\qquad\qquad \left. + [\mathrm{Re} \,
a_\ell^M(\omega)-|a_\ell^M(\omega)|^2]\right\}.
\end{eqnarray}

From the $S$ matrix elements, we can also precisely describe the
resonant behavior of the sphere as well as the geometrical and
diffractive aspects of the scattering process. Here the dual
structure of the $S$ matrix plays a crucial role. Indeed, the $S$
matrix is a function of both the frequency $\omega$ and the angular
momentum index $\ell$. It can be analytically extended into the
complex $\omega$-plane as well as into the complex $\lambda$-plane
or CAM plane. (Here $\lambda$ denotes the complex angular momentum
index replacing $\ell +1/2$ where $\ell$ is the ordinary momentum
index\cite{New82,Nus92,Grandy}.) The poles of the $S$ matrix lying
in the fourth quadrant of the complex $\omega$-plane are the complex
frequencies of the resonant modes. These resonances are determined
by solving
\begin{equation}\label{det}
D^{E,M}_\ell(\omega)=0 \quad \mathrm{for} \quad \ell =1,2,3, \dots
\end{equation}
The solutions of (\ref{det}) are denoted by $\omega_{\ell
p}=\omega^{(0)}_{\ell p}-i\Gamma _{\ell p}/2$ where
$\omega^{(0)}_{\ell p}>0$ and $\Gamma _{\ell p}>0$, the index $p$
permitting us to distinguish between the different roots of
(\ref{det}) for a given $\ell$. In the immediate neighborhood of the
resonance $\omega_{\ell p}$, $S^{E,M}_\ell(\omega)$ has the
Breit-Wigner form, i.e., is proportional to
\begin{equation}\label{BW}
\frac{\Gamma _{\ell p}/2}{\omega -\omega^{(0)}_{\ell p}+i\Gamma
_{\ell p}/2}.
\end{equation}
The structure of the $S$ matrix in the complex $\lambda$-plane
allows us, by using integration contour deformations, Cauchy's
Theorem and asymptotic analysis, to provide a semiclassical
description of scattering in terms of rays (geometrical and
diffracted). In that context, the poles of the $S$-matrix lying in
the CAM plane (the so-called Regge poles) are associated with
diffraction. They are determined by solving
\begin{equation}\label{RP}
D^{E,M}_{\lambda-1/2} ( \omega)=0 \quad \mathrm{for} \quad \omega >
0.
\end{equation}
Of course, when a connection between these two faces of the $S$
matrix can be established, resonance aspects are then
semiclassically interpreted.

In the following, we shall present some numerical results. We have
chosen to restrict ourselves to particular configurations, i.e., to
particular values of the parameters $\epsilon_h$, $\epsilon_\infty$,
$\omega_p$, $\omega_Ta/c$, $\omega_La/c$ and $\gamma$. These values
are physically realistic or, more precisely, of the same order than
physically realistic values (see, for example,
Refs.~\onlinecite{FuchsKliewer1968,AshcroftMermin}). In fact, even
for different configurations, the results we have numerically
obtained and that we shall discuss in this paper remain valid.

\subsection{Sphere with a Drude-like behavior}

\begin{figure}
\includegraphics[height=7cm,width=8.6cm]{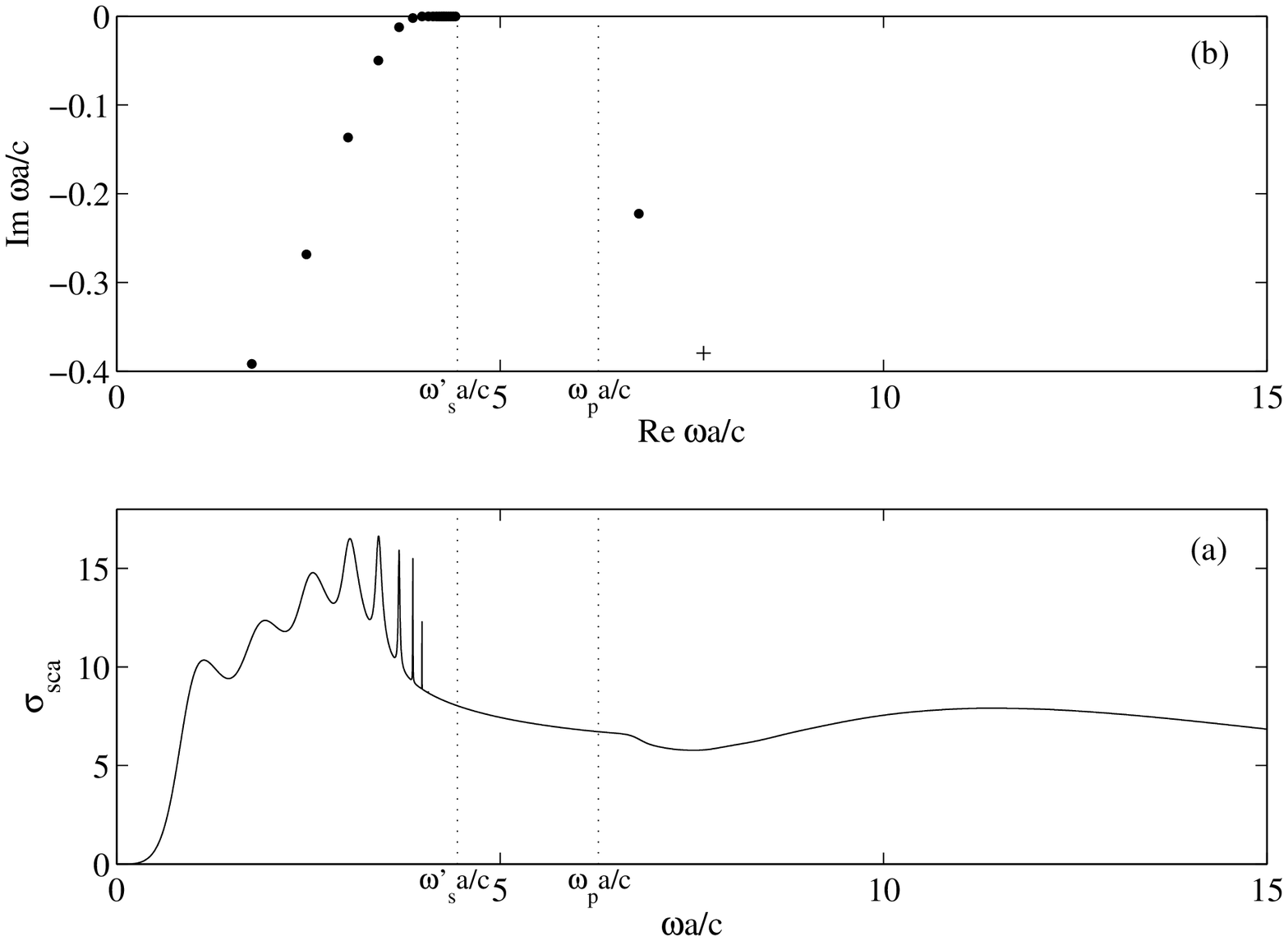}
\caption{\label{fig:crossMet_Nabs} a) Scattering cross section
$\sigma _\mathrm{sca}$. b) Scattering resonances in the complex
$\omega a/c$-plane. We consider a non-absorbing sphere:
$\epsilon_c(\omega)$ has the Drude type behavior with
$\epsilon_\infty=1$, $\omega_pa/c=2\pi$ and $\gamma =0$ while
$\epsilon_h=1$.
 Dots $(\cdot)$ correspond to poles of $S^E(\omega)$ while plus ($+$)
correspond to poles of $S^M(\omega)$.}
\end{figure}

\begin{figure}
\includegraphics[height=7cm,width=8.6cm]{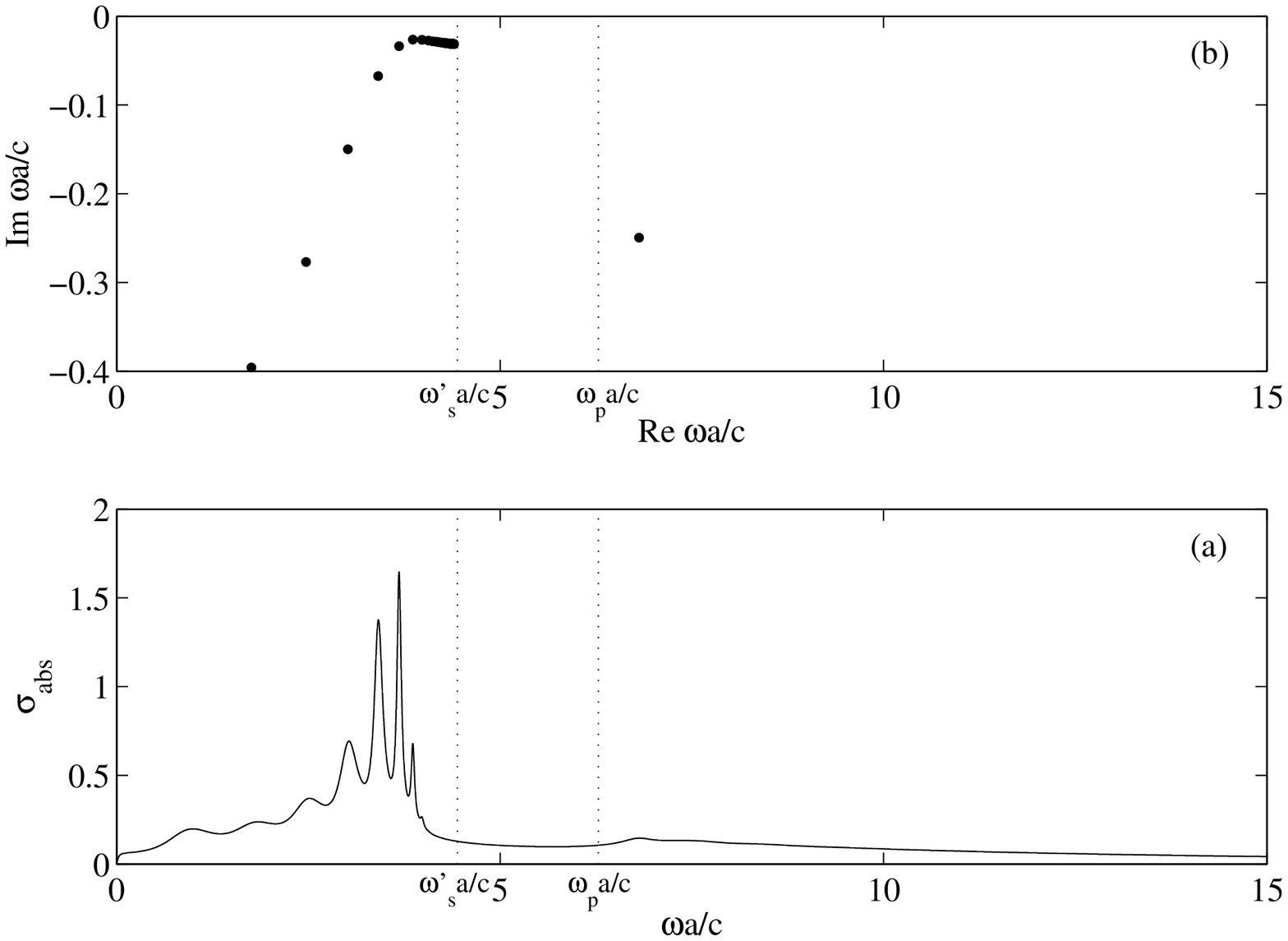}
\caption{\label{fig:crossMet_Abs} a) Absorption cross section
$\sigma _\mathrm{abs}$. b) Scattering resonances in the complex
$\omega a/c$-plane. We consider an absorbing sphere:
$\epsilon_c(\omega)$ has the Drude type behavior with
$\epsilon_\infty=1$, $\omega_pa/c=2 \pi$ and $\gamma =1/100$ while
$\epsilon_h=1$. Dots $(\cdot)$ correspond to poles of $S^E(\omega)$
while plus ($+$) correspond to poles of $S^M(\omega)$.}
\end{figure}

In Figs.~\ref{fig:crossMet_Nabs} and \ref{fig:crossMet_Abs}, we
consider the resonant aspects of a sphere embedded in vacuum
($\epsilon_h=1$) and we assume that its dielectric function presents
the Drude-like behavior given by Eq.~(\ref{PermDrude}) with
$\epsilon_\infty=1$ and $\omega_pa/c=2\pi$. We examine both the
non-absorbing case with $\gamma=0$ in Fig.~\ref{fig:crossMet_Nabs}
and the absorbing case with $\gamma=1/100$ in
Fig.~\ref{fig:crossMet_Abs}. In the non-absorbing case, we display
the scattering cross section in Fig.~\ref{fig:crossMet_Nabs}a and,
in the absorbing case, we display the absorption cross section in
Fig.~\ref{fig:crossMet_Abs}a. These cross sections are both plotted
as functions of the reduced frequency $\omega a /c$. On the two
figures, rapid variations of sharp characteristic shapes can be
observed. This strongly fluctuating behavior is due to scattering
resonances: when a pole of the $S$ matrix is sufficiently close to
the real axis in the complex $\omega$-plane, it has a strong
influence on the cross section [see Eq.~(\ref{BW})]. In
Figs.~\ref{fig:crossMet_Nabs}b and \ref{fig:crossMet_Abs}b,
resonances are exhibited for the two configurations previously
considered. A one-to-one correspondence between the peaks of the
cross sections and the resonances near the real $\omega a/c$-axis
can be clearly observed in certain frequency ranges.

More precisely and more generally, for the dielectric function
(\ref{PermDrude}) there exists in the frequency range where
$\epsilon'_c (\omega) <0$ (i.e., where $\omega \lesssim \omega_p$ if
we neglect terms in $\gamma^2$) a family of resonances associated
with $S^E$ (TM resonances). They are close to the real axis of the
complex $\omega$-plane and they converge, for large $\ell$, to the
limiting complex frequency $\omega_s$ satisfying
\begin{equation}\label{accFREQmsc_1}
\epsilon_c(\omega_s) + \epsilon_h =0.
\end{equation}
The real and imaginary parts of $\omega_s$ are easily found
perturbatively by inserting
\begin{equation}\label{accFREQmsc_2}
\omega_s = \omega'_s + i \omega''_s
\end{equation}
into (\ref{accFREQmsc_1}) and by taking into account
(\ref{FuncDiel_RetI}) with $|\epsilon''_c (\omega)| \ll |\epsilon'_c
(\omega)|$. By assuming that $|\omega''_s| \ll \omega'_s$ and by
using a first-order Taylor series expansion of
$\epsilon_c(\omega_s)$, we find that $\omega'_s$ must satisfy
\begin{subequations}
\begin{equation}\label{accFREQmsc_3}
\epsilon'_c(\omega_s') + \epsilon_h =0.
\end{equation}
and that
\begin{equation}\label{accFREQmsc_4}
\omega''_s = -\left.  \frac{\epsilon''_c(\omega) }{d \ \mathrm{Re}\,
\epsilon'_c (\omega )
  /d\omega } \right|_{\omega =\omega'_s}
\end{equation}
\end{subequations}
By using Eqs.~(\ref{FuncDiel_RetI_Met_a}) and
(\ref{FuncDiel_RetI_Met_b}) and by neglecting terms in $\gamma^2$,
we then obtain
\begin{subequations}
\begin{eqnarray}\label{accFREQmsc_5}
&  &  \omega'_s \approx \frac{\omega_p}{ \sqrt{1+ \epsilon_h /
\epsilon_\infty}}  \label{accFREQmsc_5a} \\
&  &  \omega''_s \approx -\frac{\gamma \omega_p}{2}.
\label{accFREQmsc_5b}
\end{eqnarray}
\end{subequations}

The general formulas (\ref{accFREQmsc_5a}) and (\ref{accFREQmsc_5b})
describe very well the accumulation of resonances observed in
Figs.~\ref{fig:crossMet_Nabs}b and \ref{fig:crossMet_Abs}b in the
frequency range where $\epsilon'_c(\omega)<0$. It is important to
note the existence of a shift in the imaginary part of the resonance
spectrum [see Eq.~(\ref{accFREQmsc_5b})]. It is well highlighted in
Fig.~\ref{fig:crossMet_Abs}b. It is associated with absorption and
proportional to the phenomenological damping factor $\gamma$. At
first sight, it can appear surprising that it does not depend on the
dielectric constants $\epsilon_\infty$ and $\epsilon_h$. In fact,
such a dependence only appears by working at higher orders in the
perturbative expansion used.


The resonances we have just considered are associated with the
electric part $S^E$ of the $S$ matrix and they correspond to TM
resonant modes of the sphere whose excitation frequencies belong to
the frequency range in which $\epsilon'_c (\omega) <0$. From now on
(i.e., in Secs. III and IV), we shall more particularly focus our
attention on the physical interpretation of these resonant modes. We
shall prove that the corresponding resonances are generated by an
exponentially attenuated SP propagating close to the sphere surface,
this fact justifying the term RSPM's used to denote the associated
resonant modes. Furthermore, we shall show that the SP is the
analog, in the large radius limit, to that supported by the plane
interface. Of course, there also exists, in the whole frequency
range, other families of resonances with ``higher" imaginary parts
and which are associated with both the electric and the magnetic
parts of the $S$ matrix. They correspond to the excitation of TM and
TE resonant modes of the sphere. From a physical point of view,
these modes are less interesting due to their shorter lifetime.
Indeed, in the new field of plasmonics, those are the SP's with long
propagation lengths and therefore very small attenuations that are
especially interesting from the point of view of practical
applications. Furthermore, if we consider the system photon-sphere
as an artificial atom for which the photon plays the usual role of
the electron (a point of view we shall push farther in Secs. III and
IV), we must then keep in mind that, in the scattering of a photon
with frequency $\omega^{(0)}_{\ell p}$, a decaying state (i.e., a
quasibound state) of the photon-sphere system is formed. It has a
finite lifetime proportional to $1/\Gamma _{\ell p}$. The resonant
states whose complex frequencies belong to the family generated by
the SP are therefore the most interesting because they are very
long-lived states.

\subsection{Sphere with an ionic crystal behavior}

\begin{figure}
\includegraphics[height=7cm,width=8.6cm]{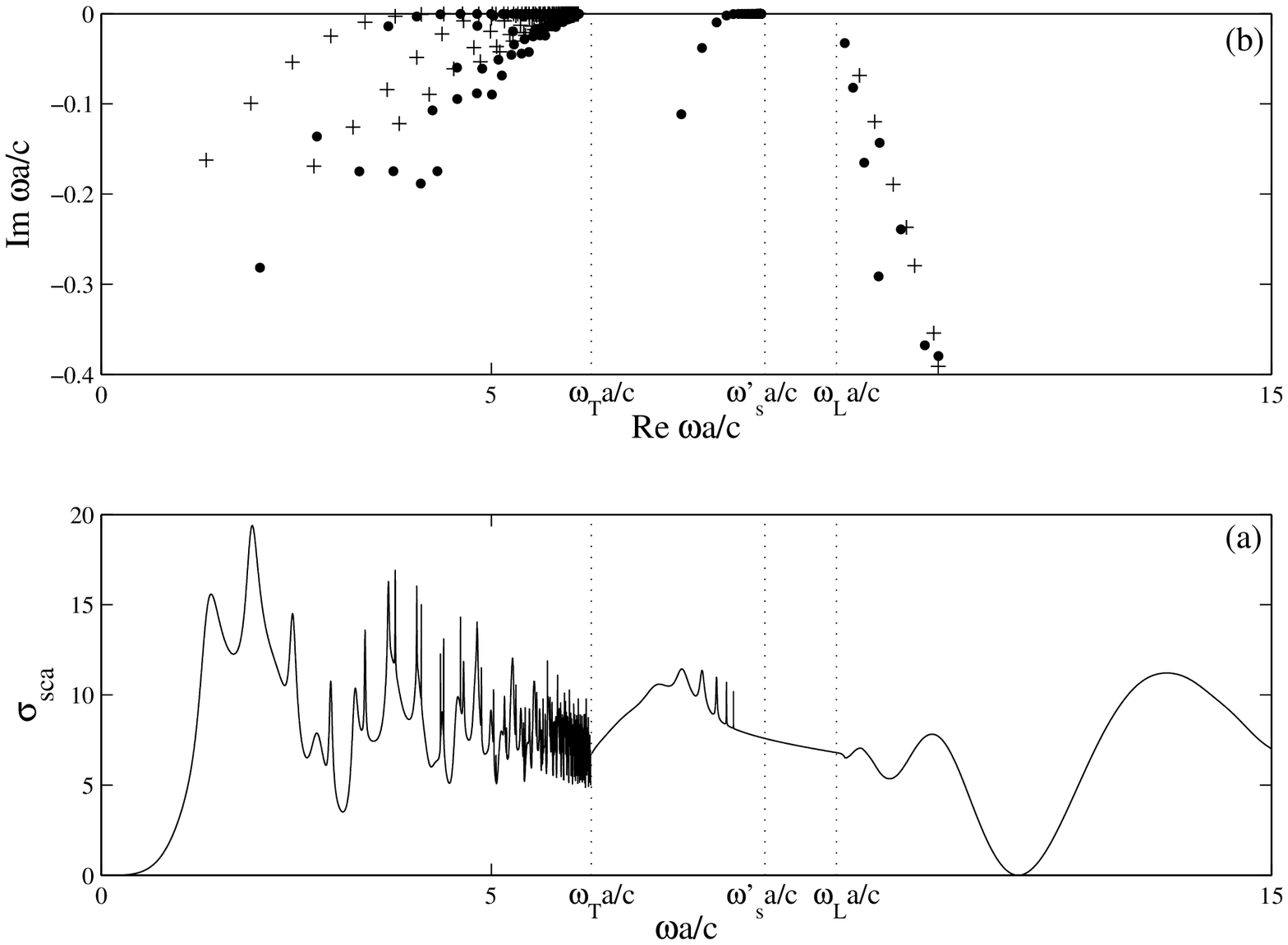}
\caption{\label{fig:crossSC_Nabs} a) Scattering cross section
$\sigma _\mathrm{sca}$. b) Scattering resonances in the complex
$\omega a/c$-plane. We consider a non-absorbing sphere:
$\epsilon_c(\omega)$ has the ionic crystal behavior with
$\epsilon_\infty=2$, $\omega_Ta/c=2\pi$, $\omega_La/c=3\pi$ and
$\gamma =0$ while $\epsilon_h=1$.
 Dots $(\cdot)$ correspond to poles of $S^E(\omega)$ while plus ($+$)
correspond to poles of $S^M(\omega)$.}
\end{figure}

\begin{figure}
\includegraphics[height=7cm,width=8.6cm]{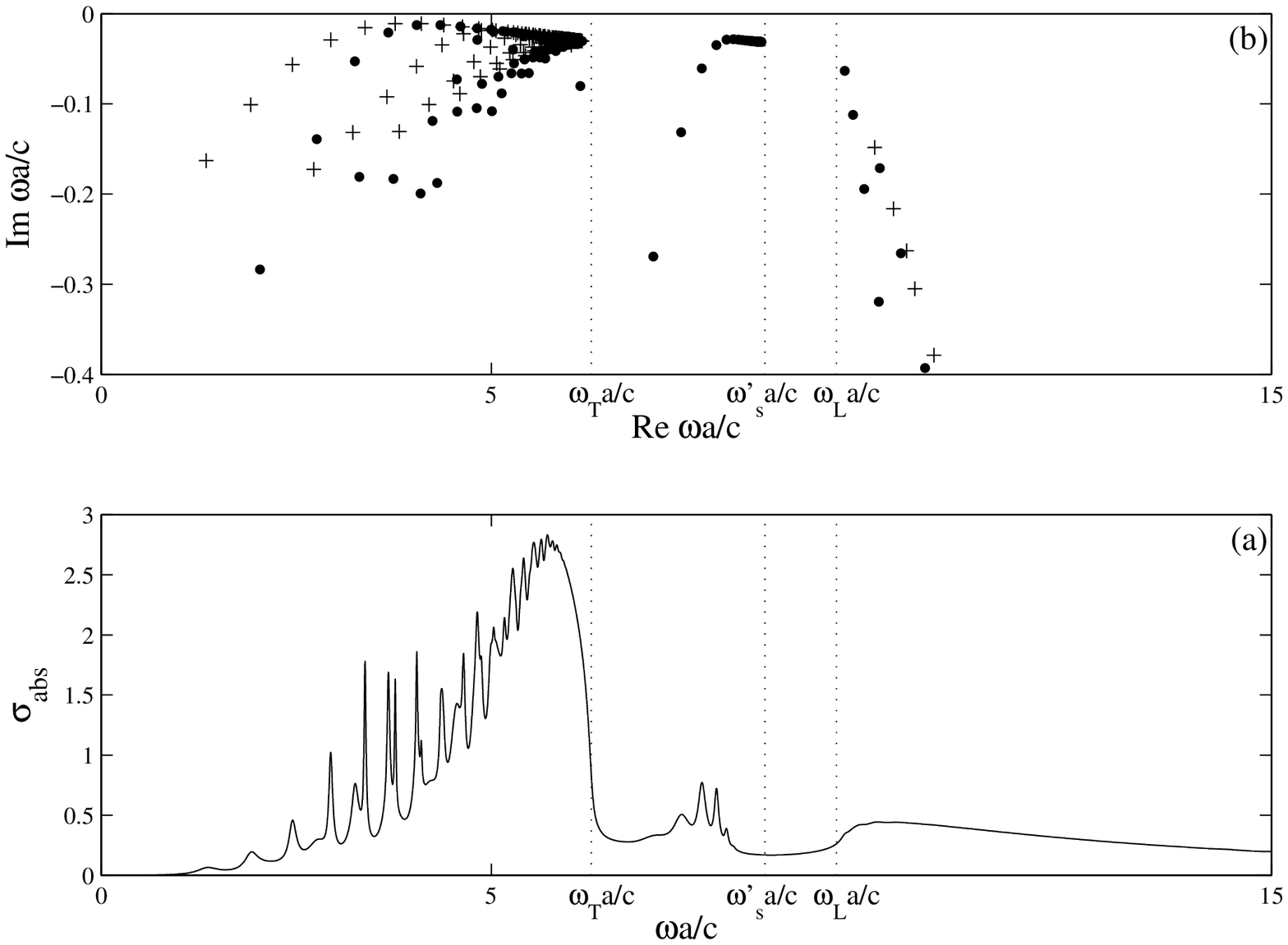}
\caption{\label{fig:crossSC_Abs} a) Absorption cross section $\sigma
_\mathrm{abs}$. b) Scattering resonances in the complex $\omega
a/c$-plane. We consider an absorbing sphere: $\epsilon_c(\omega)$
has the ionic crystal behavior with $\epsilon_\infty=2$,
$\omega_Ta/c=2\pi$, $\omega_La/c=3\pi$ and $\gamma =1/100$ while
$\epsilon_h=1$. Dots $(\cdot)$ correspond to poles of $S^E(\omega)$
while plus ($+$) correspond to poles of $S^M(\omega)$.}
\end{figure}

In Figs.~\ref{fig:crossSC_Nabs} and \ref{fig:crossSC_Abs}, we
consider the resonant aspects of a sphere embedded in vacuum
($\epsilon_h=1$) and we assume that its dielectric function presents
the ionic crystal behavior given by Eq.~(\ref{PermCristIon}) with
$\epsilon_\infty=2$, $\omega_Ta/c=2\pi$ and $\omega_La/c=3\pi$. We
examine both the non-absorbing case with $\gamma=0$ in
Fig.~\ref{fig:crossSC_Nabs} and the absorbing case with
$\gamma=1/100$ in Fig.~\ref{fig:crossSC_Abs}. In the non-absorbing
case, we display the scattering cross section in
Fig.~\ref{fig:crossSC_Nabs}a and, in the absorbing case, we display
the absorption cross section in Fig.~\ref{fig:crossSC_Abs}a. In
Figs.~\ref{fig:crossSC_Nabs}b and \ref{fig:crossSC_Abs}b, resonances
are exhibited for the two configurations. A one-to-one
correspondence between the peaks of the cross sections and the
resonances near the real $\omega a/c$-axis can be clearly observed
in certain frequency ranges.

Here again, we focus our attention on the resonances existing in the
frequency range where $\epsilon'_c (\omega) <0$ (i.e., where
$\omega_T \lesssim \omega \lesssim \omega_L$ if we neglect terms in
$\gamma^2$). There is a family of resonances associated with $S^E$
(TM resonances) close to the real axis of the complex
$\omega$-plane. They converge, for large $\ell$, to the limiting
complex frequency $\omega_s$ still satisfying
Eq.~(\ref{accFREQmsc_1}) with $\epsilon_c$ now given by
Eq.~(\ref{PermCristIon}). The real and imaginary parts of $\omega_s
= \omega'_s + i \omega''_s$ can be obtained perturbatively and we
have
\begin{subequations}
\begin{eqnarray}\label{accFREQmsc_6} &  &  \omega'_s \approx
\sqrt{\frac{\omega_L^2 + (\epsilon_h / \epsilon_\infty)
\omega_T^2 }{ 1+ \epsilon_h / \epsilon_\infty }} \label{accFREQmsc_6a} \\
&  &  \omega''_s \approx -\frac{\gamma \omega_T}{2}.
\label{accFREQmsc_6b}
\end{eqnarray}
\end{subequations}

The general formulas (\ref{accFREQmsc_6a}) and (\ref{accFREQmsc_6b})
describe very well the accumulation of resonances observed in
Figs.~\ref{fig:crossSC_Nabs}b and \ref{fig:crossSC_Abs}b in the
frequency range where $\epsilon'_c(\omega)<0$ as well as the shift
in the imaginary part of the resonance spectrum.


In Secs. III and IV, we shall provide a physical interpretation of
the family of resonances we have just considered and we shall prove
that these resonances are generated by an exponentially attenuated
SP propagating close to the sphere surface and analog, in the large
radius limit, to that supported by the plane interface. Of course,
there also exists other families of resonances but we shall not
focus our attention on them even if they seem to us physically
interesting due to their rather ``low" imaginary parts (see the
frequency range $\omega a/c \lesssim \omega_T a/c$ in
Figs.~\ref{fig:crossSC_Nabs}b and \ref{fig:crossSC_Abs}b). In fact,
it is not possible to interpret them in term of SP's analog to that
supported by the plane interface because, in frequency ranges where
$\epsilon'_c(\omega)>0$, no surface wave can be supported by the
plane interface. We think however that a semiclassical description
could be achieved (maybe in terms of whispering-gallery-type surface
waves) but it is out of the scope of the present work.

\subsection{Comparison between cylinders and spheres}

We conclude this section by making a brief comparison with the
results obtained in our previous study concerning metallic and
semiconducting cylinders\cite{ADFG_1}. Of course, in
Ref.~\onlinecite{ADFG_1} we have only considered non-absorbing
cylinders. As a consequence, a comparison between cylinders and
spheres can be achieved only in this restricted context and we can
then notice that the cross sections for the cylinders and the
spheres as well as the spectra of resonances are rather similar.
However, it should be noted that in the scattering by a sphere both
the TM and TE polarizations contribute to the cross section (we
recall that for the cylinder, the two polarizations can be studied
separately) and it is important to note that, for both scatterers,
SP's and their associated RSPM's correspond to only one polarization
(the TE polarization for the cylinder and the TM polarization for
the sphere). Of course, these last results remain valid even in the
presence of absorption.

\section{Semiclassical analysis: From the SP Regge pole to
the complex frequencies of RSPM's}

As already mentioned in Sec. II (see also
Refs.~\onlinecite{New82,Nus92,Grandy}), in the CAM approach, Regge
poles determined by solving Eq.~(\ref{RP}) are crucial to describe
diffraction as well as resonance phenomenons in terms of surface
waves. From the Regge trajectory associated with the SP supported by
the metallic or semiconducting sphere, i.e., from the curve
$\lambda_\mathrm{SP} =\lambda_\mathrm{SP} (\omega)$ traced out in
the CAM plane by the corresponding Regge pole as a function of the
frequency, we can more particularly deduce:

\quad (i) the dispersion relation
\begin{equation}\label{WNSP}
k_\mathrm{SP} (\omega) = \frac{\mathrm{Re} \, \lambda_\mathrm{SP}
(\omega)}{a}
\end{equation}
of the SP which connects its wave number $k_\mathrm{SP}$ with the
frequency $\omega$,

\quad (ii) the damping $\mathrm{Im} \, \lambda_\mathrm{SP} (\omega)$
of the SP,

\quad (iii) the phase velocity $v_p$ as well as the group velocity
$v_g$ of the SP given by
\begin{equation}\label{VpGg}
v_p = \frac{\omega a}{\mathrm{Re} \, \lambda_\mathrm{SP} (\omega)}
\quad \mathrm{and} \quad v_g = \frac{d~\omega a}{d~\mathrm{Re} \,
\lambda_\mathrm{SP} (\omega)},
\end{equation}

 \quad (iv) the semiclassical formula (a Bohr-Sommerfeld type quantization condition)
 which provides the location of the excitation frequencies
$\omega^{(0)}_{\ell \mathrm{SP}}$ of the resonances generated by the
SP:
\begin{equation}\label{sc1}
\mathrm{Re} \,  \lambda_\mathrm{SP} (\omega^{(0)}_{\ell \mathrm{SP}}
)= \ell + 1/2  \qquad \ell =1,2,\dots,
\end{equation}

\quad (v) the semiclassical formula which provides the widths of
these resonances
\begin{equation}\label{sc2} \frac{\Gamma _{\ell
\mathrm{SP}}}{2}= \left.  \frac{\mathrm{Im} \, \lambda_\mathrm{SP}
(\omega )(d \mathrm{Re} \, \lambda_\mathrm{SP} (\omega )
/d\omega)}{(d \mathrm{Re} \, \lambda_\mathrm{SP} (\omega )
/d\omega)^2 + (d \mathrm{Im} \, \lambda_\mathrm{SP} (\omega )
/d\omega)^2 } \right|_{\omega =\omega^{(0)}_{\ell \mathrm{SP}}}
\end{equation}
and which reduces to
\begin{equation}\label{sc4}
\frac{\Gamma _{\ell \mathrm{SP}}}{2}= \left.  \frac{\mathrm{Im} \,
\lambda_\mathrm{SP} (\omega )}{d \ \mathrm{Re} \,
\lambda_\mathrm{SP} (\omega ) /d\omega } \right|_{\omega
=\omega^{(0)}_{\ell \mathrm{SP}}}
\end{equation}
in the frequency range where the condition $| d  \mathrm{Re} \,
\lambda_\mathrm{SP} (\omega ) /d\omega | \gg |d  \mathrm{Im} \,
\lambda_\mathrm{SP} (\omega ) /d\omega |$ is satisfied.
 \noindent All these results can be established
by generalizing, {\it mutatis mutandis}, our approach and our
calculations developed in Refs.\onlinecite{ADFG_1,ADFG_2} for
dispersive cylinders. The transition from the dimension 2 to the
dimension 3 induces some additional technical difficulties
(vectorial treatment, existence of a caustic, asymptotics for
spherical harmonics ...) which can be rather easily overcome
following and extending the works of Newton in quantum mechanics
(see Ch. 13 of Ref.~\onlinecite{New82}) and the works of
Nussenzveig\cite{Nus92} and Grandy\cite{Grandy} in electromagnetism
of ordinary dielectric media.

\begin{figure}
\includegraphics[height=6cm,width=8.6cm]{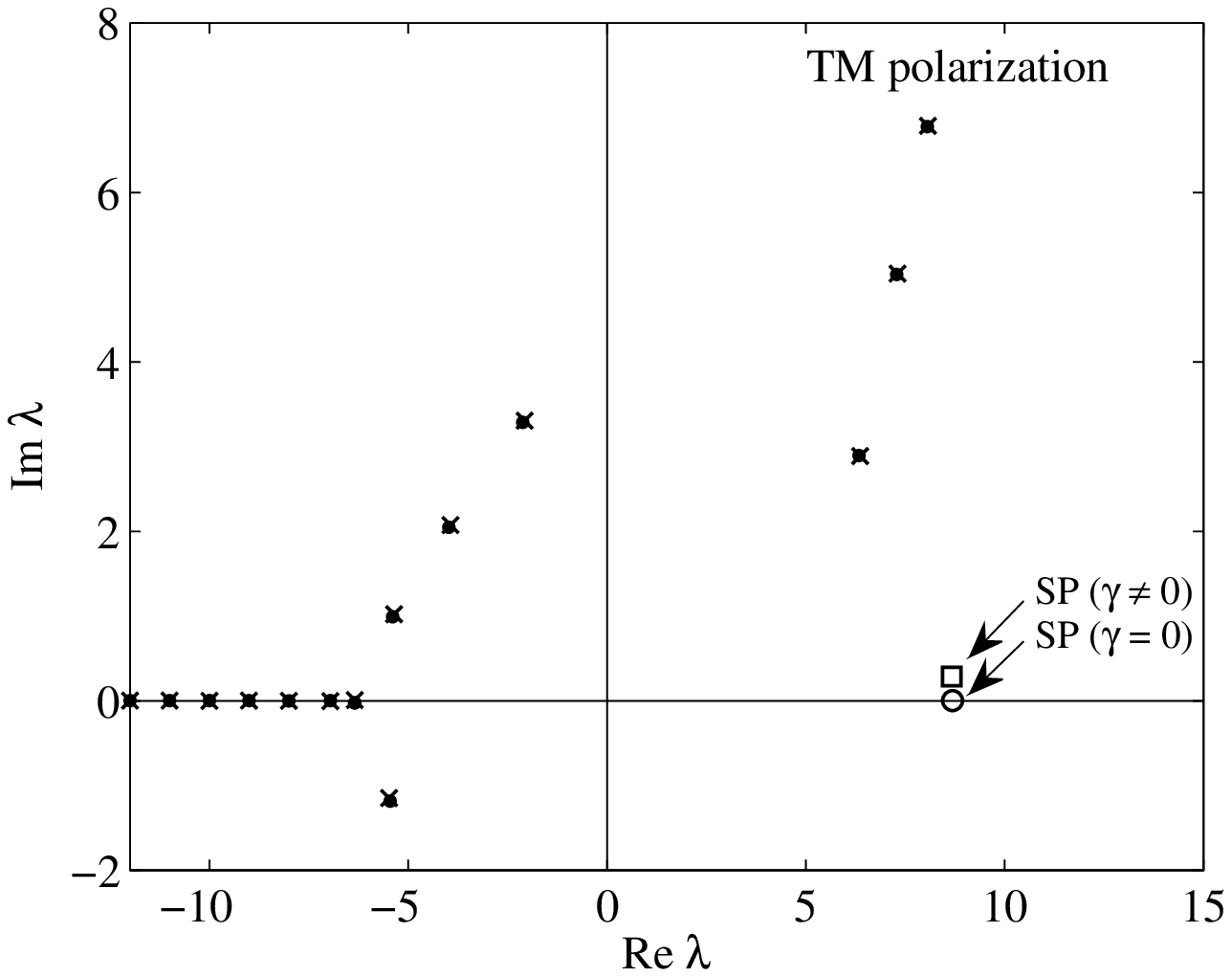}
\caption{\label{fig:RP1} Regge poles in the complex angular momentum
plane. $\epsilon_c(\omega)$ has the Drude type behavior with
$\epsilon_\infty=1$ and $\omega_pa/c=2\pi$ while $\epsilon_h=1$. The
distribution corresponds to $\omega a/c= 4$ and we have
$\epsilon'_c(\omega)<0$.  Dots $(\cdot)$ and crosses ($\mathrm{x}$)
correspond respectively to Regge poles for $\gamma=0$ and for
$\gamma=1/100$.}
\end{figure}

\begin{figure}
\includegraphics[height=6cm,width=8.6cm]{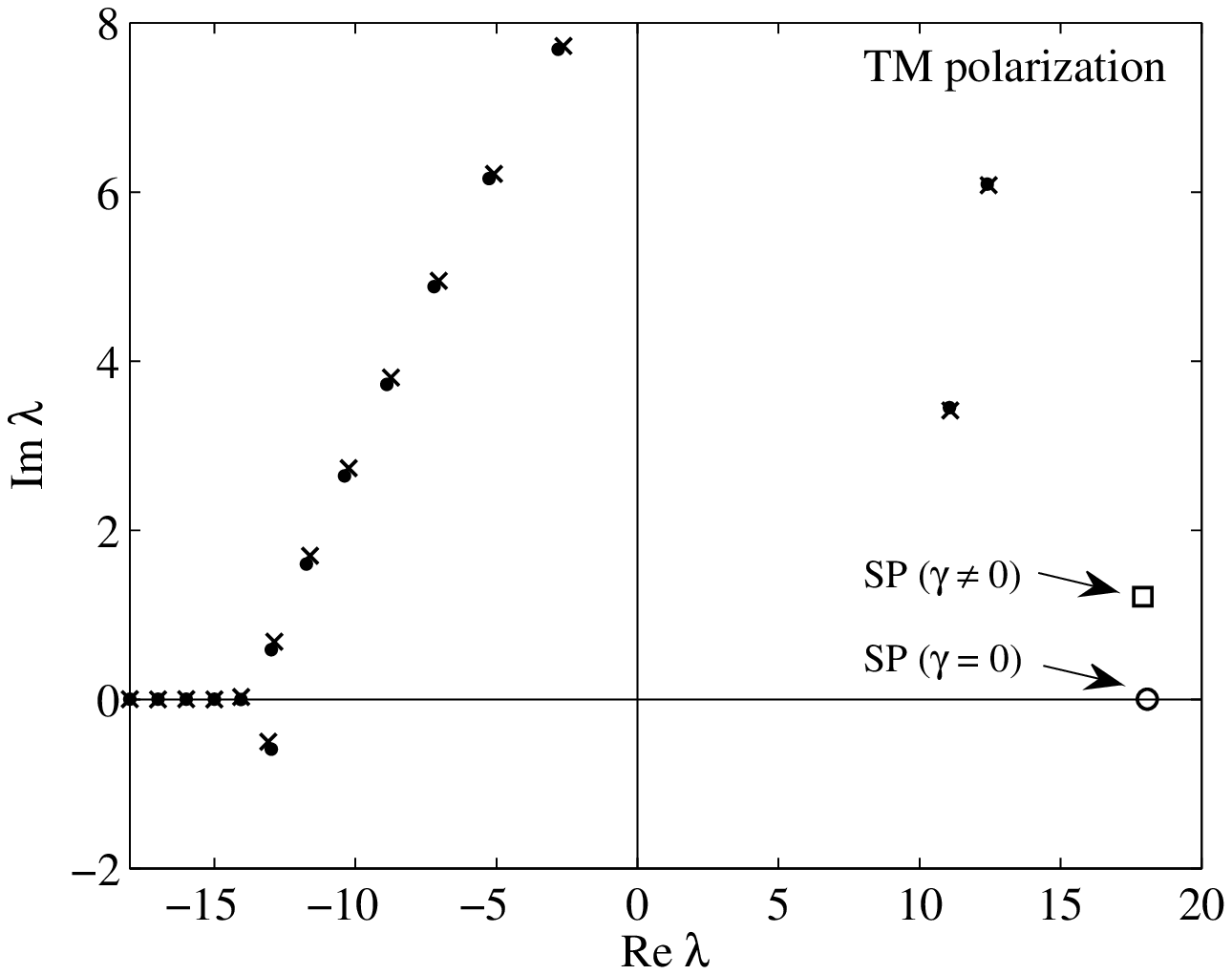}
\caption{\label{fig:RP2} Regge poles in the complex angular momentum
plane. $\epsilon_c(\omega)$ has the ionic crystal behavior with
$\epsilon_\infty=2$, $\omega_Ta/c=2\pi$ and $\omega_La/c=3\pi$ while
$\epsilon_h=1$. The distribution corresponds to $\omega a/c=8.3$ and
we have $\epsilon'_c(\omega)<0$. Dots $(\cdot)$ and crosses
($\mathrm{x}$) correspond respectively to Regge poles for $\gamma=0$
and for $\gamma=1/100$.}
\end{figure}

Figs.~\ref{fig:RP1} and \ref{fig:RP2} exhibit the distribution of
Regge poles for a sphere embedded in vacuum when
$\epsilon'_c(\omega)<0$. We only consider the Regge poles of the
electric part $S^E$ of the $S$ matrix (TM polarization) for the
configurations numerically studied in Sec. II. These Regge pole
distributions are rather similar to the distributions associated
with the ordinary dielectric sphere\cite{Nus92,Grandy}. However,
something new occurs: there exists a well-identified particular
Regge pole lying in the first quadrant of the $\lambda$-plane and
close to the real axis. This new Regge pole
$\lambda_\mathrm{SP}(\omega)$ is associated with the SP orbiting
around the metallic or semiconducting sphere. Absorption induces an
important modification of its imaginary part while it leaves
unchanged the position of the other Regge poles. For other
configurations (i.e., for other values of the parameters
$\epsilon_\infty$, $\epsilon_h$, $\omega_p$, $\omega_T$, $\omega_L$
and $\gamma$), the Regge pole distributions are not globally
different from those of Figs.~\ref{fig:RP1} and \ref{fig:RP2}. The
SP Regge pole $\lambda_\mathrm{SP}(\omega)$ is still present. By
contrast, when $\epsilon'_c(\omega)>0$ or when we consider the TE
polarization, the SP Regge pole never exists.

\begin{figure}
\includegraphics[height=6cm,width=8.6cm]{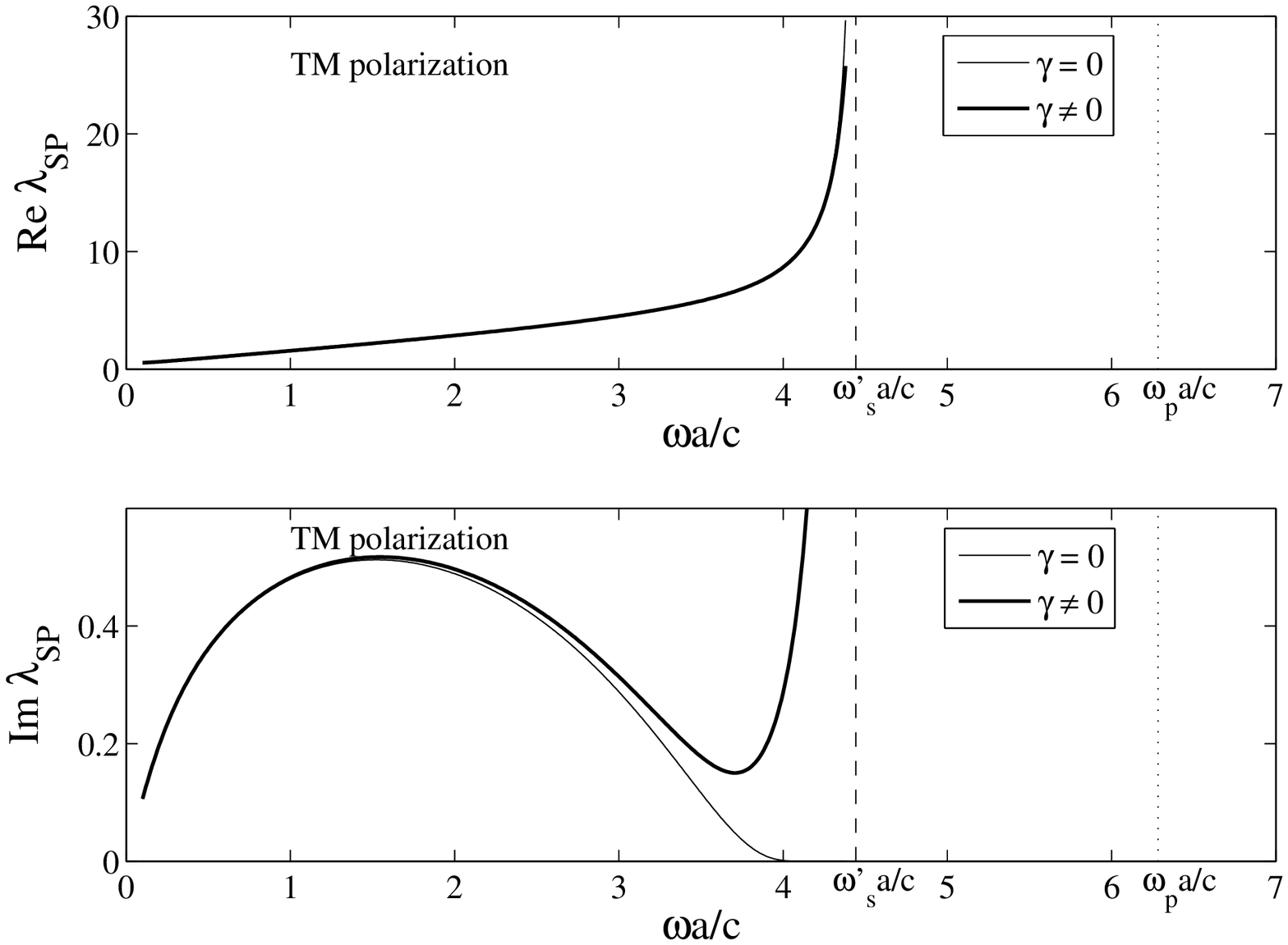}
\caption{\label{fig:RT1} Regge trajectory for the SP Regge pole:
comparison between a non-absorbing and an absorbing sphere.
$\epsilon_c(\omega)$ has the Drude type behavior with
$\epsilon_\infty=1$ and $\omega_pa/c=2\pi$ while $\epsilon_h=1$. As
$\omega a/c \to \omega'_s a/c$, the real part of the SP Regge pole
always increases indefinitely  while its imaginary part vanishes for
$\gamma=0$ and increases indefinitely for $\gamma \not= 0$.}
\end{figure}
\begin{figure}
\includegraphics[height=6cm,width=8.6cm]{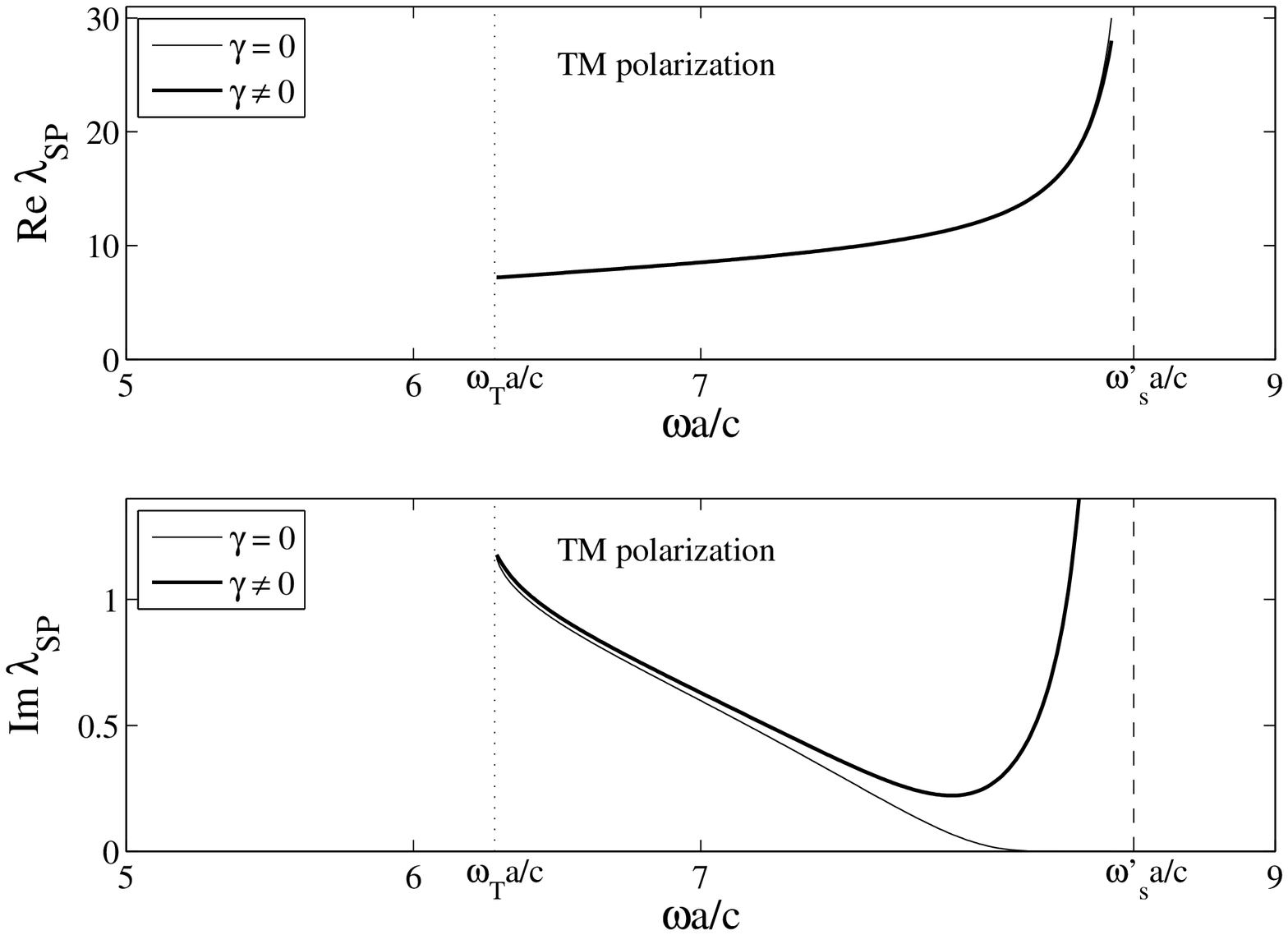}
\caption{\label{fig:RT2} Regge trajectory for the SP Regge pole:
comparison between a non-absorbing and an absorbing sphere.
$\epsilon_c(\omega)$ has the ionic crystal behavior with
$\epsilon_\infty=2$, $\omega_Ta/c=2\pi$ and $\omega_La/c=3\pi$ while
$\epsilon_h=1$. As $\omega a/c \to \omega'_s a/c$, the real part of
the SP Regge pole increases indefinitely while its imaginary part
vanishes for $\gamma=0$ and increases indefinitely for $\gamma \not=
0$.}
\end{figure}

In Figs.~\ref{fig:RT1} and \ref{fig:RT2}, we have displayed the
Regge trajectories of the SP for the two configurations previously
studied. Absorption does not modify the global behavior of the real
part of the SP Regge pole. It should be also noted that, as $\omega
\to \omega'_s $, this real part increases indefinitely. By contrast,
absorption increases significantly the imaginary part of the Regge
trajectory of the SP: for the non-absorbing sphere, this imaginary
part vanishes as $\omega \to \omega'_s $ while, for the absorbing
sphere, it first exhibits a minimum and then increases indefinitely
as $\omega \to \omega'_s $. In other words, absorption has a strong
influence on the damping of the SP but does not modify its
dispersion relation. For other configurations (i.e., for other
values of the parameters $\epsilon_\infty$, $\epsilon_h$,
$\omega_p$, $\omega_T$, $\omega_L$ and $\gamma$), the SP Regge pole
behavior is very similar. The minimum of the imaginary part of the
Regge trajectory always exists (for the absorbing sphere). This
feature could be interesting with in mind practical applications
using real materials.

\begin{table}
\caption{\label{tab:table1} The first complex frequencies of RSPM's
(TM polarization). $\epsilon_c(\omega)$ has the Drude type behavior
with $\epsilon_\infty=1$, $\omega_p a/c=2\pi$ and $\gamma=0$ while
$\epsilon_h=1$.}
\begin{ruledtabular}
\begin{tabular}{ccccc}
        &\quad Exact \quad& \quad
Exact \quad&  Semiclassical  &  Semiclassical  \\
 $\ell $   &$   (\omega ^{(0)}_{\ell {\mathrm{SP}}})a/c  $
 &  $  (\Gamma _{\ell {\mathrm{SP}}}/2)a/c  $
 & $  (\omega ^{(0)}_{\ell {\mathrm{SP}}})a/c   $ &  $ (\Gamma _{\ell {\mathrm{SP}}}/2)a/c  $  \\
\hline
1&  0.897849     &  0.407409   &  0.944599 &  0.381875  \\
2&  1.762396   &    0.391747   &  1.734451 &  0.380517 \\
3&  2.475717     &  0.268405  &   2.430280 &  0.267887  \\
4&  3.019638     &  0.136587  &   2.992907 &  0.138470  \\
5&  3.412422    &   0.049796  &   3.404834 &  0.050338  \\
6&  3.683304   &    0.012179  &   3.682351 &  0.012207  \\
7&  3.862800     &  0.001935  &   3.862752 &  0.001934  \\
8&  3.982206   &    0.000209  &   3.982204 &  0.000208   \\
9&  4.065198     &  0.000016  &   4.065196 &  0.000016  \\
10& 4.125633      & 0.000001   &  4.125631 &  0.000001
\end{tabular}
\end{ruledtabular}
\end{table}
\begin{table}
\caption{\label{tab:table2} The first complex frequencies of RSPM's
(TM polarization). $\epsilon_c(\omega)$ has the Drude type behavior
with $\epsilon_\infty=1$, $\omega_p a/c=2\pi$ and $\gamma=1/100$
while $\epsilon_h=1$.}
\begin{ruledtabular}
\begin{tabular}{ccccc}
        &\quad Exact \quad& \quad
Exact \quad&  Semiclassical  &  Semiclassical  \\
 $\ell $   &$   (\omega ^{(0)}_{\ell {\mathrm{SP}}})a/c  $
 &  $  (\Gamma _{\ell {\mathrm{SP}}}/2)a/c  $
 & $  (\omega ^{(0)}_{\ell {\mathrm{SP}}})a/c   $ &  $ (\Gamma _{\ell {\mathrm{SP}}}/2)a/c  $  \\
\hline
1&  0.894546     &  0.408326   &  0.942572 &  0.382753  \\
2&  1.757428   &    0.395822   &  1.731152 &  0.384335 \\
3&  2.470440     &  0.276889  &   2.426006 &  0.276611  \\
4&  3.015603     &  0.149847  &   2.988477 &  0.152411  \\
5&  3.410258    &   0.067443  &   3.401552 &  0.068144  \\
6&  3.682529   &    0.033581  &   3.681107 &  0.033585  \\
7&  3.862589     &  0.026173  &   3.863243 &  0.026087  \\
8&  3.982111   &    0.026321  &   3.983622 &  0.026089   \\
9&  4.065111     &  0.027345  &   4.067129 &  0.027396  \\
10& 4.125541      & 0.028160   &  4.128020 &  0.028289
\end{tabular}
\end{ruledtabular}
\end{table}

Tables~\ref{tab:table1}, \ref{tab:table2}, \ref{tab:table3} and
\ref{tab:table4} present samples of complex frequencies of RSPM's
for the two configurations previously considered. They are
calculated from the semiclassical formulas (\ref{sc1}) and
(\ref{sc2}) by using the Regge trajectories numerically determined
by solving (\ref{RP}) (see Figs.~\ref{fig:RT1} and \ref{fig:RT2}).
We can observe a very good agreement between the ``exact" and the
semiclassical spectra for ``high" frequencies as well as a rather
good agreement for ``low" frequencies. Furthermore, from the
behavior of Regge trajectories near the limiting frequencies
$\omega'_s$ and the semiclassical formulas (\ref{sc1}) and
(\ref{sc2}), we easily obtain the existence of the families of
resonances close to the real axis of the complex $\omega$-plane
which converge, for large $\ell$, to the limiting frequency
$\omega'_s+i\omega''_s$. In conclusion, we have established a
connection between the complex frequencies of RSPM's and a
particular surface wave, the so-called SP, described by a particular
Regge pole of the electric part of the $S$ matrix and which orbits
around the sphere.

We conclude this section by making a brief comparison with the
results obtained in our previous study concerning metallic and
semiconducting cylinders\cite{ADFG_1}. Of course, such a comparison
can be achieved only in the non-absorbing case. From Regge
trajectories (see Figs.~6 and 7 of Ref.~\onlinecite{ADFG_1} and
Figs.~\ref{fig:RT1} and \ref{fig:RT2} of the present article), we
can observe that the behavior of the SP orbiting around a
metallic/semiconducting sphere is, at first sight, rather similar to
the behavior of the SP orbiting around a metallic/semiconducting
cylinder, even if they correspond to different polarizations (TE
polarization for the cylinder and TM polarization for the sphere).
In fact, as we shall see in the next section, in the absence of
absorption the transition from two dimensions to three dimensions
induces some curvature corrections on the wave numbers
$k_\mathrm{SP} (\omega)$ of the SP's and the behaviors are identical
only in the radius limit $a\to \infty$, i.e., in the flat interface
limit. In the presence of absorption, we shall prove than a
supplementary correction associated with the imaginary part of the
complex dielectric constant is necessary in order to interpret the
behavior observed in Figs.~\ref{fig:RT1} and \ref{fig:RT2}.

\begin{table}
\caption{\label{tab:table3} Some complex frequencies of RSPM's (TM
polarization). $\epsilon_c(\omega)$ has the ionic crystal behavior
with $\epsilon_\infty=2$, $\omega_Ta/c=2\pi$, $\omega_La/c=3\pi$ and
$\gamma=0$ while $\epsilon_h=1$.}
\begin{ruledtabular}
\begin{tabular}{ccccc}
        &\quad Exact \quad& \quad
Exact \quad&  Semiclassical  &  Semiclassical  \\
 $\ell $   &$   (\omega ^{(0)}_{\ell {\mathrm{SP}}})a/c $
 &  $  (\Gamma _{\ell {\mathrm{SP}}}/2)a/c  $
 & $  (\omega ^{(0)}_{\ell {\mathrm{SP}}})a/c   $ &  $ (\Gamma _{\ell {\mathrm{SP}}}/2)a/c  $  \\
\hline
9&  7.436965     &  0.111396   & 7.405909    &  0.115670  \\
10&  7.700893    &   0.037894  & 7.694243    &  0.038461  \\
11& 7.887534     &  0.009632   &  7.886710   &  0.009713 \\
12& 8.015334      & 0.001839    & 8.015239    & 0.001844  \\
13& 8.104140      & 0.000274   &  8.104087    & 0.000273  \\
14& 8.168509      & 0.000034    & 8.168477    & 0.000034  \\
15& 8.217128      & 0.000005    & 8.217069    & 0.000003  \\
16& 8.255038     &  0.000002   &  8.254956    & 0.000000  \\
17& 8.285324     &  0.000001   &  8.285231    & 0.000000
\end{tabular}
\end{ruledtabular}
\end{table}
\begin{table}
\caption{\label{tab:table4} Some complex frequencies of RSPM's (TM
polarization). $\epsilon_c(\omega)$ has the ionic crystal behavior
with $\epsilon_\infty=2$, $\omega_Ta/c=2\pi$, $\omega_La/c=3\pi$ and
$\gamma=1/100$ while $\epsilon_h=1$.}
\begin{ruledtabular}
\begin{tabular}{ccccc}
        &\quad Exact \quad& \quad
Exact \quad&  Semiclassical  &  Semiclassical  \\
 $\ell $   &$   (\omega ^{(0)}_{\ell {\mathrm{SP}}})a/c $
 &  $  (\Gamma _{\ell {\mathrm{SP}}}/2)a/c  $
 & $  (\omega ^{(0)}_{\ell {\mathrm{SP}}})a/c   $ &  $ (\Gamma _{\ell {\mathrm{SP}}}/2)a/c  $  \\
\hline
9&  7.433460     &  0.131422   & 7.399537    &  0.137680  \\
10&  7.699297    &   0.060583  & 7.690606    &  0.061609  \\
11& 7.887001     &  0.034650   &  7.885688   &  0.035096 \\
12& 8.015186      & 0.028573    & 8.016067    & 0.028737  \\
13& 8.104082      & 0.028141   & 8.105881    & 0.028098  \\
14& 8.168462      & 0.028649    & 8.170808    & 0.027735  \\
15& 8.217080      & 0.029138    & 8.219861    & 0.029527  \\
16& 8.254988     &  0.029514   & 8.258154    & 0.029563  \\
17& 8.285273     &  0.029801   & 8.288906    & 0.029931
\end{tabular}
\end{ruledtabular}
\end{table}

\section{Semiclassical analysis: Asymptotics for
the SP and physical description}

An analytical expression for the Regge pole $\lambda_\mathrm{SP}$
and therefore a deeper physical understanding of the SP behavior can
be obtained by solving Eq.~(\ref{RP}) for $\lambda =
\lambda_\mathrm{SP}$. For the TM polarization, Eq.~(\ref{RP})
reduces to [see Eq.~(\ref{SE3b})]
\begin{equation} \label{RPSP1}
 \frac{\sqrt{\epsilon_c(\omega)}}{\sqrt{\epsilon_h}} \frac{\zeta
 _{\lambda_\mathrm{SP}-1/2
}^{(1)^{\prime }}\left( \sqrt{\epsilon_h} a\omega/c\right) }{\zeta
_{\lambda_\mathrm{SP}-1/2 }^{(1)} \left( \sqrt{\epsilon_h}
a\omega/c\right)}=\frac{\psi _{\lambda_\mathrm{SP}-1/2}^{\prime
}\left(\sqrt{\epsilon_c(\omega)} a\omega/c\right)}{\psi
_{\lambda_\mathrm{SP}-1/2 }\left(\sqrt{\epsilon_c(\omega)}
a\omega/c\right)}.
\end{equation}
This equation cannot be solved exactly but only perturbatively. With
this aim in view, we must first replace Ricatti-Bessel functions by
spherical Bessel functions. We then use their relations with the
ordinary Bessel functions (see Ref.~\onlinecite{AS65})
\begin{equation} \label{SpB-B}
j_\lambda (z)=\sqrt{\frac{\pi}{2z}}J_{\lambda+1/2}(z) \quad
\mathrm{and} \quad h^{(1)}_\lambda
(z)=\sqrt{\frac{\pi}{2z}}H^{(1)}_{\lambda+1/2}(z)
\end{equation}
and we finally replace the Bessel function $J_\lambda(z)$ by the
modified Bessel function $I_\lambda(z)$ (see Ref.~\onlinecite{AS65})
in order to take into account the fact that $\mathrm{Re} \,
\epsilon_c(\omega) <0$. Eq.~(\ref{RPSP1}) then reduces to
\begin{eqnarray} \label{RPSP2}
& & \frac{1}{\sqrt{\epsilon_h}}
\frac{H_{\lambda_\mathrm{SP}}^{(1)'}(\sqrt{\epsilon_h} \omega a/c
)}{H_{\lambda_\mathrm{SP}}^{(1)}(\sqrt{\epsilon_h} \omega a/c )}
+\frac{1}{2\epsilon_h}\left(\frac{c}{\omega a} \right) \nonumber \\
& & \quad = - \frac{1}{\sqrt{-\epsilon_c(\omega)}}
\frac{I'_{\lambda_\mathrm{SP}}(\sqrt{-\epsilon_c(\omega)} \omega a/c
)}{I_{\lambda_\mathrm{SP}}(\sqrt{-\epsilon_c(\omega)} \omega a/c
)}+\frac{1}{2\epsilon_c(\omega) }\left(\frac{c}{\omega a} \right).
\nonumber \\ & &
\end{eqnarray}
This equation must be compared with Eq.~(26) of
Ref.~\onlinecite{ADFG_1} which provides the SP Regge pole for the
cylinder. The first term on the left-hand side and the right-hand
side of Eq.~(\ref{RPSP2}) are exactly those appearing in Eq.~(26) of
Ref.~\onlinecite{ADFG_1}. The two others terms are simple curvature
corrections due to the change of dimension. As a consequence,
Eq.~(\ref{RPSP2}) can be solved following the method used in order
to solve Eq.~(26) of Ref.~\onlinecite{ADFG_1} but, now, it is
important to take carefully into account the fact that the
dielectric function $\epsilon_c(\omega)$ has an imaginary part [see
Eqs.~(\ref{FuncDiel_RetI_Met_b}) and (\ref{FuncDiel_RetI_SC_b})]. It
should be noted that the existence of such curvature corrections has
been first observed by Berry\cite{BerryMV1975} in the restricted
case of non-dispersive and non-absorbing spheres.

On the right-hand side of (\ref{RPSP2}), by assuming
$|\lambda_\mathrm{SP}| \gg |\sqrt{-\epsilon_c(\omega)} \omega a/c
|$, we can write\cite{AS65,ADFG_1}
\begin{eqnarray} \label{rhsRPSP2}
&  & - \frac{1}{\sqrt{-\epsilon_c(\omega)}}
\frac{I'_{\lambda_\mathrm{SP}}(\sqrt{-\epsilon_c(\omega)} \omega a/c
)}{I_{\lambda_\mathrm{SP}}(\sqrt{-\epsilon_c(\omega)}
\omega a/c )} \nonumber \\
&  & \qquad \qquad \qquad \sim \frac{\left[ \lambda_\mathrm{SP}^2 -
\epsilon_c(\omega) (\omega a/c)^2 \right]^{1/2}}{\epsilon_c(\omega)
(\omega a/c)}.
\end{eqnarray}
On the left-hand side of (\ref{RPSP2}), by assuming
$|\lambda_\mathrm{SP}| \gg \sqrt{\epsilon_h} \omega a/c $, we can
write\cite{WatsonBessel,Nuss65,ADFG_1}
\begin{eqnarray} \label{lhsRPSP2}
&  &\frac{1}{\sqrt{\epsilon_h}}
\frac{H_{\lambda_\mathrm{SP}}^{(1)'}(\sqrt{\epsilon_h} \omega a/c
)}{H_{\lambda_\mathrm{SP}}^{(1)}(\sqrt{\epsilon_h} \omega a/c )}
\nonumber \\ &  & \qquad \sim -  \frac{\left[ \lambda_\mathrm{SP}^2
- \epsilon_h
(\omega a/c)^2 \right]^{1/2}}{\epsilon_h (\omega a/c)}  \nonumber \\
& &  \qquad  \quad \times \left( 1- i \, e^{2
\alpha(\lambda_\mathrm{SP},\sqrt{\epsilon_h} \omega a/c)} \right)
\end{eqnarray}
where
\begin{equation}
\alpha(\lambda ,z)   =  (\lambda^2 - z^2)^{1/2} -\lambda \ln \left(
\frac{\lambda + (\lambda^2 - z^2)^{1/2}}{z} \right).
\label{AsympDebyeIc}
\end{equation}
It should be noted that in Eq.~(\ref{lhsRPSP2}) we have taken into
account an exponentially small contribution (the term $\exp{[2
\alpha(\lambda_\mathrm{SP},\sqrt{\epsilon_h} \omega a/c)]}$) which
lies beyond all orders in perturbation theory. This term can be
captured by carefully taking into account Stokes phenomenon
\cite{Berry89,Dingle73,SegurTL91,BerryHowls90} and is necessary to
extract the asymptotic expression of the imaginary part of
$\lambda_\mathrm{SP}$. (In Eq.~(\ref{lhsRPSP2}) we have given to the
Stokes multiplier function the value $1/2$.) For more precision, we
refer to our previous article\cite{ADFG_1}.

By using (\ref{rhsRPSP2}) and (\ref{lhsRPSP2}) as well as
$|\epsilon''_c (\omega)| \ll |\epsilon'_c (\omega)|$,
Eq.~(\ref{RPSP2}) can be solved perturbatively and we find
\begin{widetext}
\begin{subequations}\label{RPSPasymBallO1}
\begin{equation}
\mathrm{Re} \, \lambda_\mathrm{SP} (\omega ) \sim  \left(
\frac{\omega a}{c} \right) \sqrt{  \frac{\epsilon_h
\epsilon'_c(\omega )}{\epsilon_h + \epsilon'_c(\omega )}} \left( 1+
\frac{1}{2\sqrt{-[\epsilon_h + \epsilon'_c(\omega )]}}\left(
\frac{c}{\omega a} \right) \right), \label{RPSPasymBallOa}
\end{equation}
and
\begin{equation}
\mathrm{Im} \, \lambda_{\mathrm{SP}}(\omega)= \mathrm{Im}_1 \,
\lambda_{\mathrm{SP}}(\omega) +  \mathrm{Im}_2 \,
\lambda_{\mathrm{SP}}(\omega) \label{RPSPasymBallOb}
\end{equation}
with
\begin{eqnarray}
&  & \mathrm{Im}_1 \, \lambda_{\mathrm{SP}}(\omega) \sim
\left[\frac{{\epsilon'_c}^2 (\omega ) }{ {\epsilon'_c}^2 (\omega
)-\epsilon_h^2} \right]  \frac{ \left[\mathrm{Re} \,
\lambda_{\mathrm{SP}} (\omega )\right]^2- \epsilon_h\left( {\omega
a/c} \right)^2 }{\mathrm{Re} \, \lambda_{\mathrm{SP}} (\omega )}
\left[ 1-\frac{\epsilon_h /\epsilon'_c(\omega )}{2\sqrt{-[\epsilon_h
+ \epsilon'_c(\omega )]}}\left( \frac{c}{\omega a} \right) \right]
\exp \lbrace{2 \alpha[\mathrm{Re} \, \lambda_{\mathrm{SP}} (\omega )
,\sqrt{\epsilon_h}  \omega a/c]\rbrace}, \nonumber \\
& & \label{RPSPasymBallOc}
\\
&  & \mathrm{Im}_2 \, \lambda_{\mathrm{SP}}(\omega) \sim \left(
\frac{\omega a}{c} \right) \sqrt{  \frac{\epsilon_h
\epsilon'_c(\omega )}{\epsilon_h + \epsilon'_c(\omega )}}
\frac{\epsilon_h \epsilon''_c(\omega )}{2\epsilon'_c(\omega )
[\epsilon_h + \epsilon'_c(\omega )]}   \left[ 1+
\frac{1}{2\sqrt{-[\epsilon_h + \epsilon'_c(\omega )]}}\left(
\frac{c}{\omega a} \right) \right]. \label{RPSPasymBallOd}
\end{eqnarray}
\end{subequations}
\end{widetext}

\begin{figure}
\includegraphics[height=6cm,width=8.6cm]{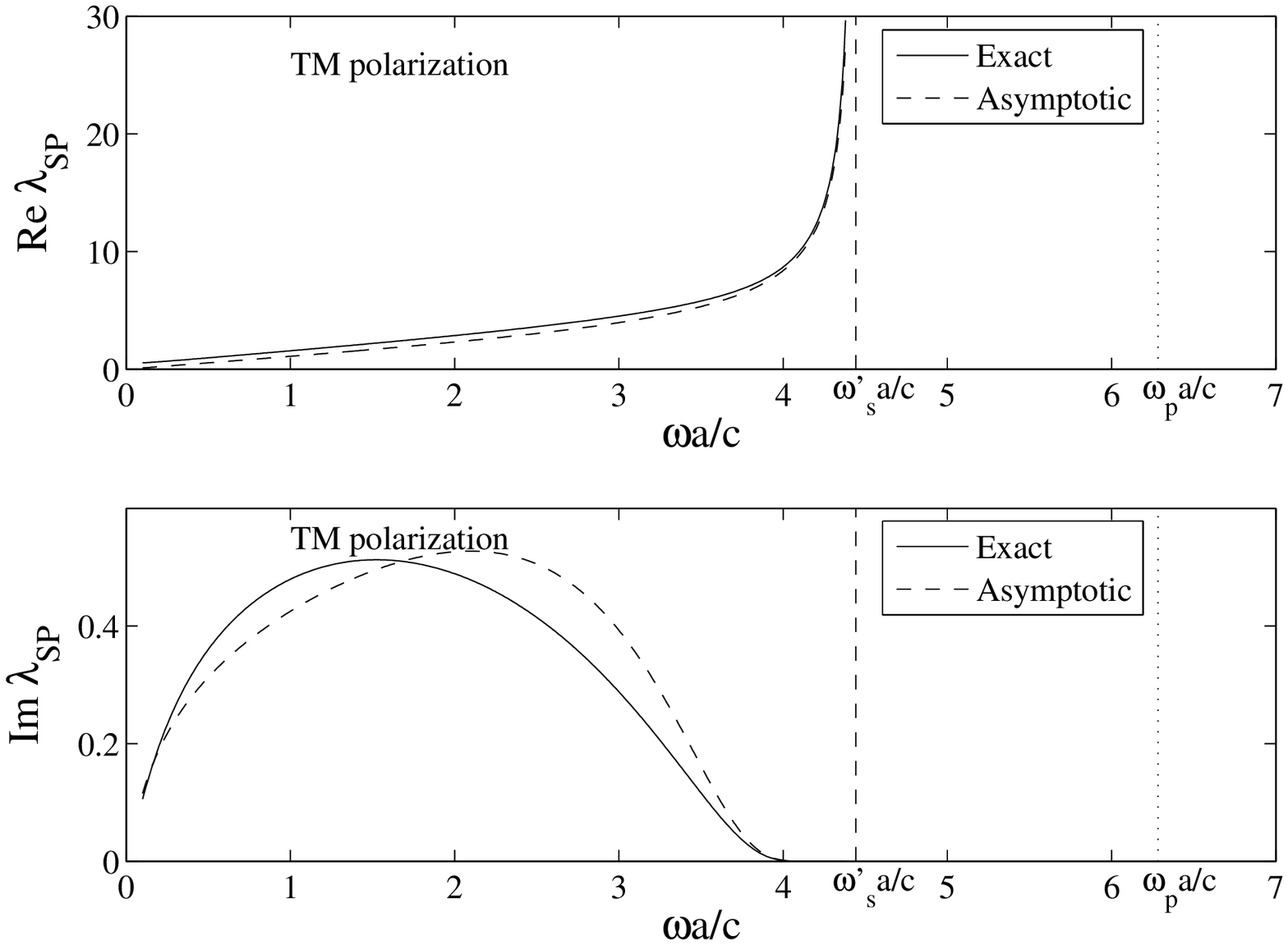}
\caption{\label{fig:RT1asymp} Regge trajectory for the SP Regge
pole. Comparison between exact and asymptotic theories.
$\epsilon_c(\omega)$ has the Drude type behavior with
$\epsilon_\infty=1$, $\omega_pa/c=2\pi$ and $\gamma=0$ while
$\epsilon_h=1$.}
\end{figure}
\begin{figure}
\includegraphics[height=6cm,width=8.6cm]{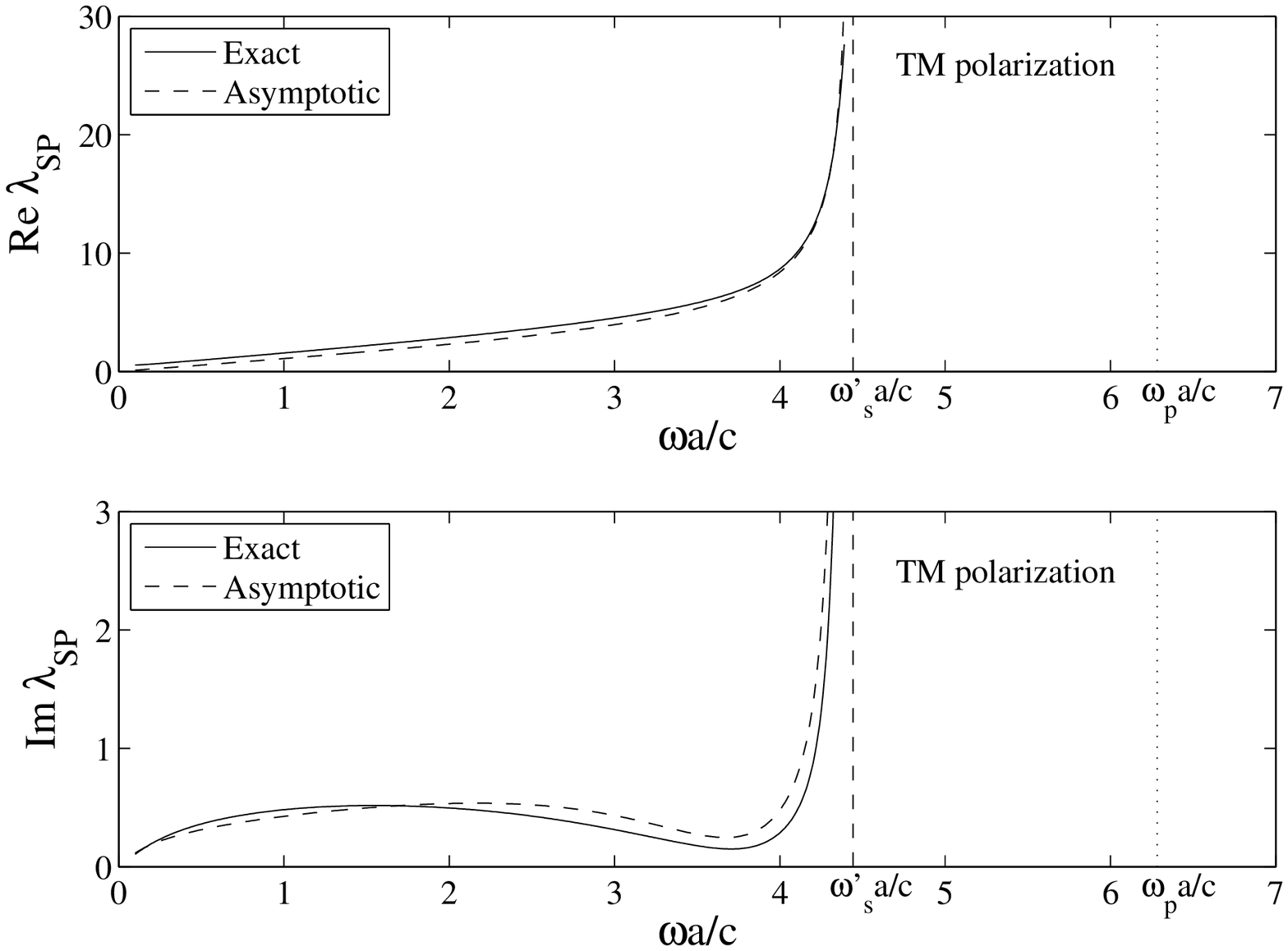}
\caption{\label{fig:RT2asymp} Regge trajectory for the SP Regge
pole. Comparison between exact and asymptotic theories.
$\epsilon_c(\omega)$ has the Drude type behavior with
$\epsilon_\infty=1$, $\omega_pa/c=2\pi$ and $\gamma=1/100$ while
$\epsilon_h=1$.}
\end{figure}
\begin{figure}
\includegraphics[height=6cm,width=8.6cm]{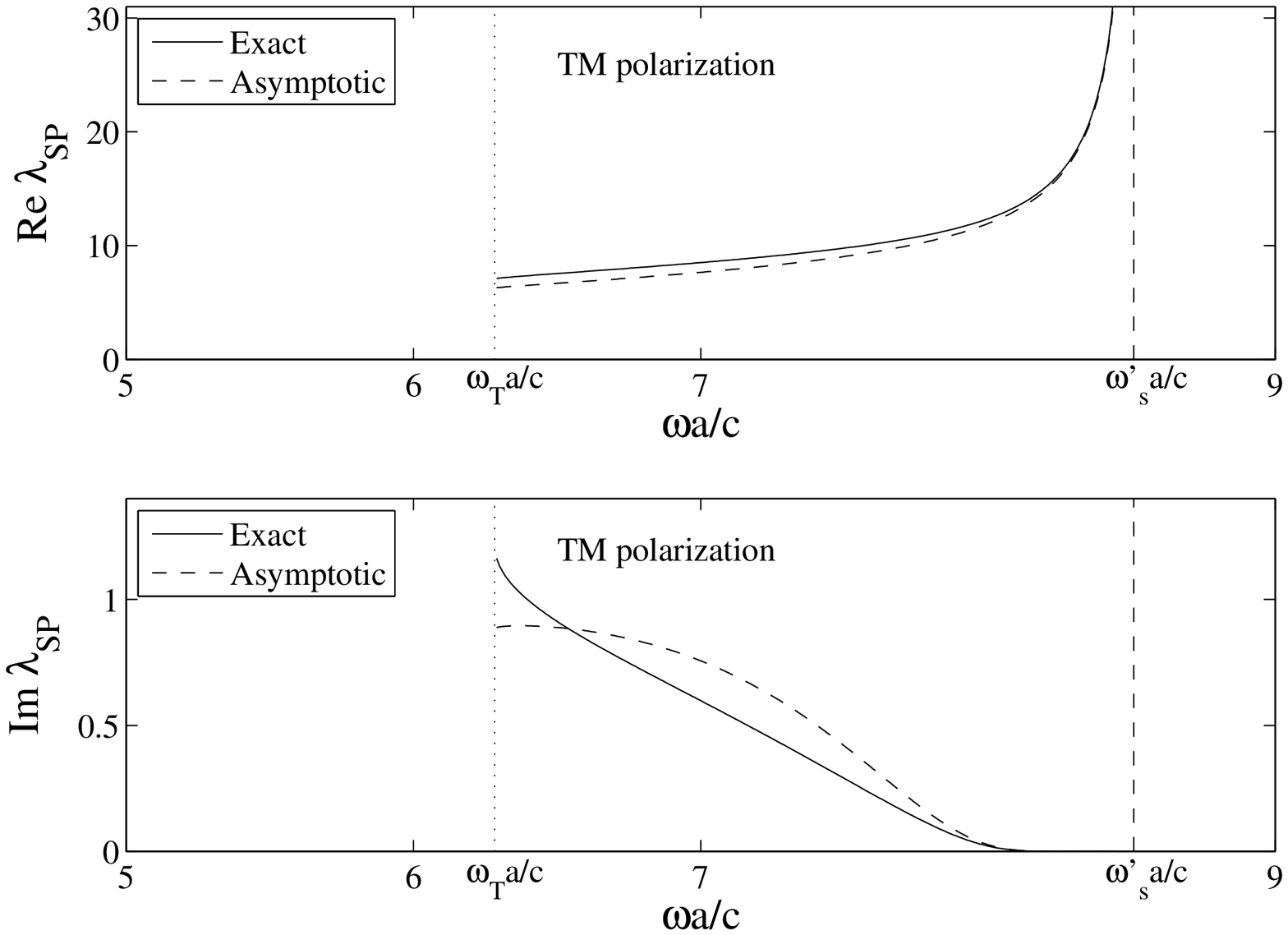}
\caption{\label{fig:RT3asymp}Regge trajectory for the SP Regge pole.
Comparison between exact and asymptotic theories.
$\epsilon_c(\omega)$ has the ionic crystal behavior with
$\epsilon_\infty=2$, $\omega_Ta/c=2\pi$, $\omega_La/c=3\pi$ and
$\gamma=0$ while $\epsilon_h=1$.}
\end{figure}
\begin{figure}
\includegraphics[height=6cm,width=8.6cm]{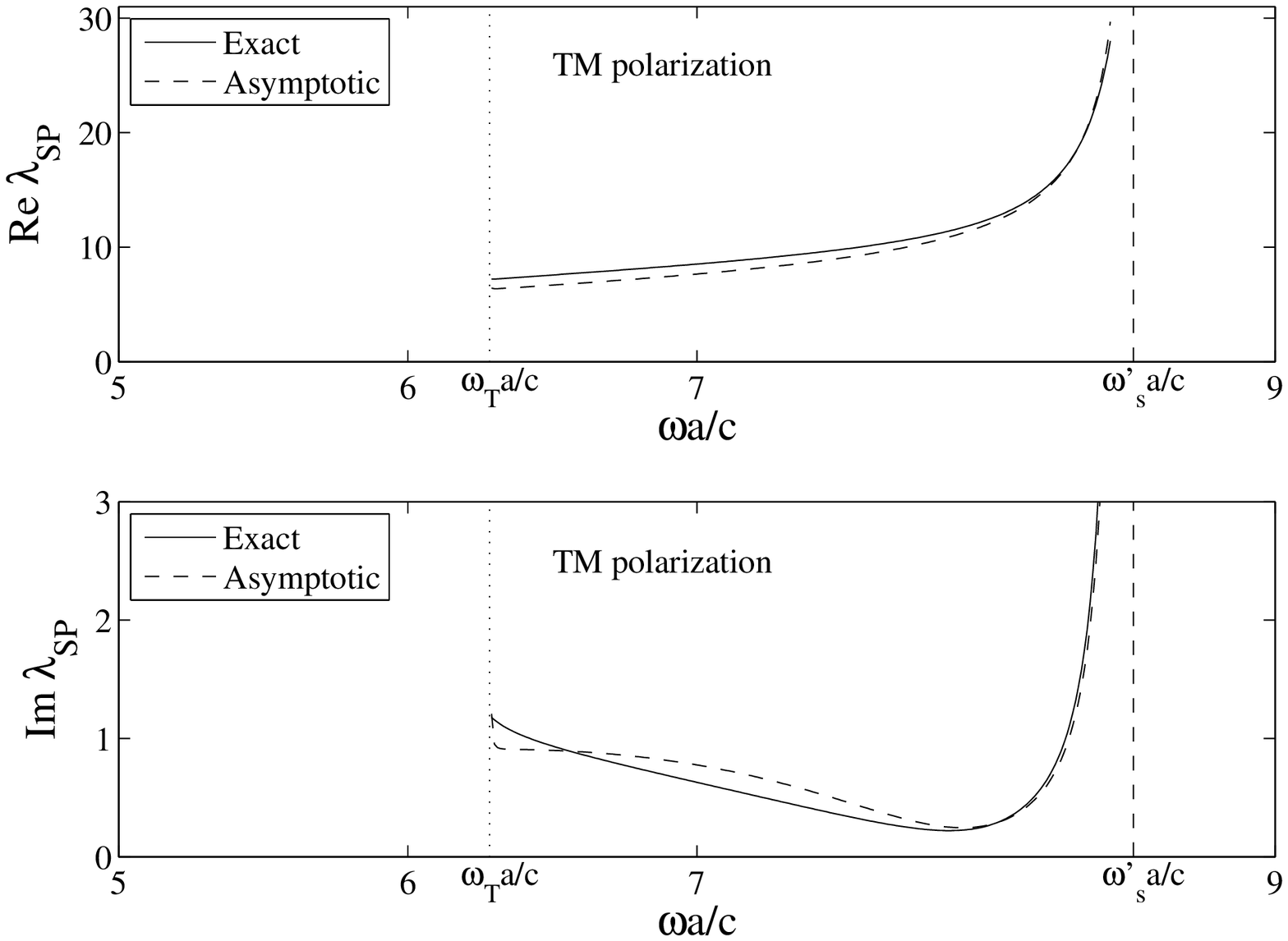}
\caption{\label{fig:RT4asymp}Regge trajectory for the SP Regge pole.
Comparison between exact and asymptotic theories.
$\epsilon_c(\omega)$ has the ionic crystal behavior with
$\epsilon_\infty=2$, $\omega_Ta/c=2\pi$, $\omega_La/c=3\pi$ and
$\gamma=1/100$  while $\epsilon_h=1$.}
\end{figure}

Equations (\ref{RPSPasymBallOa}), (\ref{RPSPasymBallOb}),
(\ref{RPSPasymBallOc}) and (\ref{RPSPasymBallOd}) provide analytic
expressions for the dispersion relation and the damping of the SP.
The following important features must be noted:

\qquad -- The SP only exists in the frequency range where
$\epsilon_h+\epsilon'_c (\omega) <0$. Its dispersion relation [see
Eqs.~(\ref{RPSPasymBallOa}) and (\ref{WNSP})] only depends on the
real part of the dielectric function: thus, it is slightly modify by
absorption. Its attenuation [see Eqs.~(\ref{RPSPasymBallOb}),
(\ref{RPSPasymBallOc}) and (\ref{RPSPasymBallOd})] is compounded by
two contributions. The first one [see Eq.~(\ref{RPSPasymBallOc}]
only depends on the real part of the dielectric function and
presents an exponentially small attenuation. It fully describes the
attenuation of the SP propagating on a non-absorbing sphere. The
second contribution [see Eq.~(\ref{RPSPasymBallOd}] is proportional
to the imaginary part of the dielectric function and is therefore
directly linked to absorption. It increases indefinitely as $\omega
\to \omega'_s$ and semiclassically explains the behavior already
described in Sec. III.

\qquad -- The wave number $k_{\mathrm{SP}} (\omega)$ associated with
the SP is obtained from (\ref{RPSPasymBallOa}) and (\ref{WNSP}) and
is given by
\begin{eqnarray}\label{WNSPinfH}
& & k_{\mathrm{SP}} (\omega) \sim \left( \frac{\omega }{c} \right)
\sqrt{  \frac{\epsilon_h \epsilon'_c(\omega )}{\epsilon_h +
\epsilon_c(\omega )}} \nonumber \\
&  & \qquad \times \left[ 1+ \frac{1}{2\sqrt{-[\epsilon_h +
\epsilon'_c(\omega )]}}\left( \frac{c}{\omega a} \right) \right].
\end{eqnarray}
This relation could permit us to derive analytically the phase
velocity $v_p= \omega / k_\mathrm{SP}(\omega )$ as well as the group
velocity $v_g= d\omega / dk_\mathrm{SP}(\omega )$ of the SP [see
also Eq.~ (\ref{VpGg})].

\qquad -- For $a \to \infty$ -- i.e., in the flat interface limit --
we recover the usual dispersion relation of a SP supported by a flat
metal-dielectric or semiconductor-dielectric interface (see, for
example, Ref.~\onlinecite{Raether88}).

\qquad -- The imaginary part (\ref{RPSPasymBallOc}) of
$\lambda_{\mathrm{SP}}$ vanishes for $a \to \infty$: the SP
supported by the flat interface between an ordinary dielectric and a
non-absorbing metallic or semiconducting medium has no damping (see,
for example, Ref.~\onlinecite{Raether88}).

 \qquad -- By inserting the expression
(\ref{RPSPasymBallOa}) for $\lambda_\mathrm{SP} (\omega )$ into the
Bohr-Sommerfeld quantization condition (\ref{sc1}), we obtain an
eighth-order polynomial equation which can be solved numerically and
which provides rather precisely the resonance excitation frequencies
$\omega^{(0)}_{\ell {\mathrm{SP}}}$.

\qquad -- We have numerically tested the formulas
(\ref{RPSPasymBallOa}), (\ref{RPSPasymBallOb}),
(\ref{RPSPasymBallOc}) and (\ref{RPSPasymBallOd}) for various values
of the parameters $\epsilon_\infty$, $\epsilon_h$, $\omega_p$,
$\omega_T$, $\omega_L$ and $\gamma$. In all cases, they provide
rather good approximations for $\mathrm {Re} ~ \lambda_\mathrm{SP}$
and $\mathrm {Im} ~ \lambda_\mathrm{SP}$ (see
Figs.~\ref{fig:RT1asymp}, \ref{fig:RT2asymp}, \ref{fig:RT3asymp} and
\ref{fig:RT4asymp} for the two configurations previously studied).

The connection with the results obtained in our previous study
concerning metallic and semiconducting cylinders\cite{ADFG_1} can be
made. The transition from the cylinder to the sphere induces some
curvature corrections on the real and imaginary parts of the Regge
pole of the SP [the second term on the right-hand side of
(\ref{RPSPasymBallOa}), the third term on the right-hand side of
(\ref{RPSPasymBallOc}) and the fourth term on the right-hand side of
(\ref{RPSPasymBallOd})] which come from the second terms on the
left-hand side and the right-hand side of Eq.~(\ref{RPSP2}). Without
the curvature corrections, Eqs.~(\ref{RPSPasymBallOa}),
(\ref{RPSPasymBallOb}), (\ref{RPSPasymBallOc}) and
(\ref{RPSPasymBallOd}) reduce to
\begin{subequations}\label{RPSPasymBallO1_cyl}
\begin{equation}
\mathrm{Re} \, \lambda_\mathrm{SP} (\omega ) \sim  \left(
\frac{\omega a}{c} \right) \sqrt{  \frac{\epsilon_h
\epsilon'_c(\omega )}{\epsilon_h + \epsilon'_c(\omega )}} ,
\label{RPSPasymBallOa_cyl}
\end{equation}
and
\begin{equation}
\mathrm{Im} \, \lambda_{\mathrm{SP}}(\omega)= \mathrm{Im}_1 \,
\lambda_{\mathrm{SP}}(\omega) +  \mathrm{Im}_2 \,
\lambda_{\mathrm{SP}}(\omega) \label{RPSPasymBallOb_cyl}
\end{equation}
with
\begin{eqnarray}
&  & \mathrm{Im}_1 \, \lambda_{\mathrm{SP}}(\omega) \sim
\left[\frac{{\epsilon'_c}^2 (\omega ) }{ {\epsilon'_c}^2 (\omega
)-\epsilon_h^2} \right]  \nonumber \\
& & \qquad \qquad \times\frac{ \left[\mathrm{Re} \,
\lambda_{\mathrm{SP}} (\omega )\right]^2- \epsilon_h\left( {\omega
a/c} \right)^2 }{\mathrm{Re} \, \lambda_{\mathrm{SP}} (\omega )}
\nonumber \\
& & \qquad \qquad \times \exp \lbrace{2 \alpha[\mathrm{Re} \,
\lambda_{\mathrm{SP}} (\omega ) ,\sqrt{\epsilon_h}  \omega
a/c]\rbrace},  \label{RPSPasymBallOc_cyl}
\\
&  & \mathrm{Im}_2 \, \lambda_{\mathrm{SP}}(\omega) \sim \left(
\frac{\omega a}{c} \right) \sqrt{  \frac{\epsilon_h
\epsilon'_c(\omega )}{\epsilon_h + \epsilon'_c(\omega )}}
\frac{\epsilon_h \epsilon''_c(\omega )}{2\epsilon'_c(\omega )
[\epsilon_h + \epsilon'_c(\omega )]}  . \nonumber
\\
& & \label{RPSPasymBallOd_cyl}
\end{eqnarray}
\end{subequations}
These last expressions are the semiclassical formulas providing the
dispersion relation and the damping of the SP propagating on the
cylinder in the non-absorbing and absorbing cases. They are much
more general that the corresponding formulas obtained in our
previous paper\cite{ADFG_1}. Indeed, Eq.~(\ref{RPSPasymBallOa_cyl})
has been already obtained in Ref.~\onlinecite{ADFG_1} in the
restricted context of the non-absorbing cylinder [see Eq.~(40a)]
and, in the same way, Eq.~(\ref{RPSPasymBallOc_cyl}) has been
obtained in a less practical form [see Eq.~(40b) in
Ref.~\onlinecite{ADFG_1}].

\section{Conclusion and perspectives}

In this article, we have introduced the CAM method in order (i) to
study the interaction of electromagnetic waves with non-absorbing or
absorbing metallic and semiconducting spheres and (ii) to completely
describe the resonant aspects of this problem. This allows us to
provide a physical explanation for the excitation mechanism of
RSPM's as well as a simple mathematical description of the surface
wave (i.e., the SP) that generates them. It should be noted that our
approach is not limited to the metals and semiconductors described
by (\ref{PermDrude}) and (\ref{PermCristIon}) but remains still
valid for more general materials (see the conclusion of
Ref.~\onlinecite{ADFG_1}).

In Ref.~\onlinecite{ADFG_1}, we have developed a new picture of the
photon-cylinder system (see also the analysis of the works of Ito
and Sakoda \cite{Sakoda2001b,Sakoda2001} in the conclusion of
Ref.~\onlinecite{ADFG_1}): it can be viewed as an artificial atom
for which the photon plays the role of an electron. {\it Mutatis
mutandis}, this picture is still valid for the photon-sphere system:
RSPM's are long-lived quasibound states for this atom, the
associated complex frequencies are Breit-Wigner-type resonances
while the trajectory of the SP which generates them and which is
supported by the sphere surface is a Bohr-Sommerfeld-type orbit.
Recently, Guzatov and Klimov \cite{GuzatovKlimov2007} have also
developed the analogy between a metallic sphere and an ordinary atom
and, more generally, between a cluster of metallic spheres and
ordinary molecules, introducing on that occasion the terms
``plasmonic atom" and ``plasmonic molecules". It should be noted
that their approach uses the quasi-static approximation for the
description of the electromagnetic field. As a consequence, they
have found that the ``energy levels" of the photon-sphere system are
real (bound states). In fact, as we have shown, the imaginary parts
of the ``energy levels" do not vanish (quasi-bound states): they
correspond to exponentially small terms lying beyond all orders in
perturbation theory.

Our results could be useful (i) in the context of three-dimensional
photonic crystal physics (the existence of dispersionless band for
the arrays of metallic or semiconducting spheres is associated with
the excitation of the sphere RPSM's), (ii) in cavity quantum
electrodynamics and, more generally, (iii) in the context of
nanotechnologies and plasmonics. Here, we shall more particularly
focus our discussion to the possible applications to the Casimir
effect.

Since the precise experiments carried out by Lamoreaux
\cite{Lamoreaux97} in 1997, it is necessary to completely describe,
from a theoretical point of view, the Casimir interaction between
two metallic spheres or between a metallic sphere and a plane (for a
review on recent experiments and problematic, we refer to
Ref.~\onlinecite{BordagETMM2001}). A semiclassical description of
this interaction could be achieved by extending to electrodynamics
the Korringa-Kohn-Rostoker (KKR) type method developed in
Ref.~\onlinecite{BMetWirzba06} for the scalar Casimir effect.
Because at short distance RSPM's provide the dominant contribution
to the Casmir interaction \cite{Genet04,Henkel04}, it would be then
necessary to carefully take into account the contributions of all
the periodic orbits associated with the SP. In this context, the
description that we gave in Sec. IV could be helpful.

\begin{acknowledgments}
We are grateful to Rosalind Fiamma for help with the English.
\end{acknowledgments}

\bibliography{SPsphereMandSC}%
\end{document}